\documentclass[iop]{emulateapj}

\shorttitle{Planet Traps and Planetary Populations in the Mass-Period Diagram}
\shortauthors{Hasegawa \& Pudritz}

\begin{document}

\title{Planetary Populations in the Mass-Period Diagram: A Statistical Treatment of Exoplanet Formation and the Role of Planet Traps}

\author{Yasuhiro Hasegawa\altaffilmark{1} and Ralph E. Pudritz\altaffilmark{2}}
\affil{Department of Physics and Astronomy, McMaster University,
    Hamilton, ON L8S 4M1, Canada}
\email{yasu@asiaa.sinica.edu.tw, pudritz@physics.mcmaster.ca}

\altaffiltext{1}{currently EACOA fellow at Institute of Astronomy and Astrophysics, Academia Sinica (ASIAA), Taipei 10641, Taiwan}

\altaffiltext{2}{Origins Institute, McMaster University, Hamilton, ON L8S 4M1, Canada}

\begin{abstract}
The rapid growth in the number of known exoplanets has revealed the existence of several distinct planetary populations in 
the observed mass-period diagram. Two of the most surprising are, (1) the concentration of gas giants around 1AU and 
(2) the accumulation of a large number of low-mass planets with tight orbits, also known as super-Earths and hot Neptunes.  
We have recently shown that protoplanetary disks have multiple planet traps that are characterized by orbital radii in the disks and 
halt rapid type I planetary migration. By coupling planet traps with the standard core accretion scenario, we showed that one can account for 
the positions of planets in the mass-period diagram. In this paper,  we demonstrate quantitatively that most gas giants formed at planet traps 
tend to end up around 1 AU with most of these being contributed by dead zones and ice lines. In addition, we show that a large fraction of 
super-Earths and hot Neptunes are formed as "failed" cores of gas giants - this population being constituted by comparable contributions 
from dead zone and heat transition traps. Our results are based on the evolution of forming planets in an ensemble of disks 
where we vary only the lifetimes of disks as well as their mass accretion rates onto the host star. We show that a statistical treatment of the evolution of 
a large population of planetary cores initially caught in planet traps accounts for the existence of three distinct exoplantary populations - the hot Jupiters, 
the more massive planets at roughly orbital radii around 1 AU orbital, and the short period SuperEarths and hot Neptunes. 
There are very few evolutionary tracks that feed into the large orbital radii characteristic of the imaged Jovian planet and this
is in accord with the result of recent surveys that find a paucity of Jovian planets beyond 10 AU.  Finally, 
we find that low-mass planets in tight orbits become the dominant planetary population for low mass stars ($M_* \le 0.7 M_{\odot}$), 
in agreement with the previous studies which show that the formation of gas giants is preferred for massive stars.
\end{abstract}

\keywords{accretion, accretion disks  --- methods: analytical --- planet-disk interactions --- planets and satellites: formation --- 
protoplanetary disks --- turbulence}

\section{Introduction} \label{intro}

Surveys for exoplanets around solar-type stars \citep[e.g.,][]{us07,mml11,bkb11} have discovered over 900 planets and several thousand
planetary candidates.   
These growing data sets increasingly reveal the existence of distinct exoplanet populations that are readily discerned in the structure of the mass-semimajor 
axis diagram \citep{cl13}.    If the planetary populations that inhabit these are related to planet 
formation processes, a central question is how do such regions become populated by planets forming in disks around their host stars?

\begin{figure}
\begin{center}
\includegraphics[width=9cm]{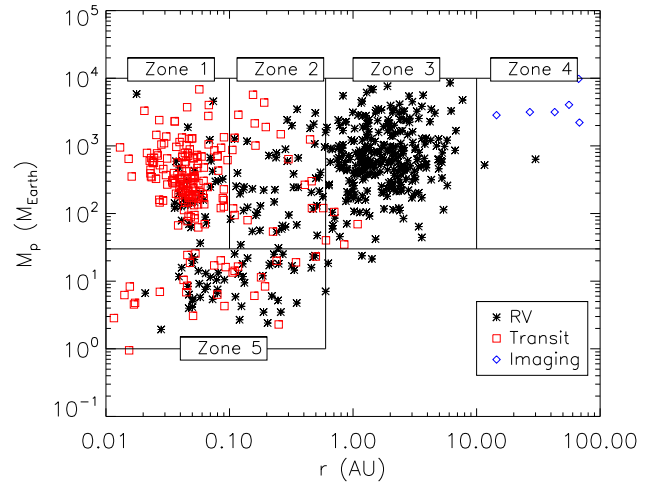}
\caption{Observed exoplanets in the mass-semimajor axis diagram. The data are taken from the website: http://exoplanet.eu/. Exoplanets 
observed by the radial velocity technique are denoted by the black stars, the transit techniques are by the red squares, and the direct imaging is 
by the blue diamonds. As suggested by \citet{cl13}, the data imply that 5 populations can be defined:  Zone 1 contains hot Jupiters, Zone 2 
the exclusion of gas giants, Zone 3 features a pile up of gas giants, Zone 4 has the distant planets, and Zone 5 contains low-mass planets with tight orbits, also known 
as super-Earths and hot Neptunes. }
\label{fig1}
\end{center}
\end{figure}

Following \citet{cl13}, we divide the mass-semimajor axis diagram into five distince regions  (see Fig. \ref{fig1}). 
The population consisting of hot Jupiters belongs to Zone 1. Hot Jupiters in Figure \ref{fig1} are significantly exaggerated by 
the transit observations done by the {\it Kepler} that mainly detect exoplanets with short orbital periods. Indeed, the radial velocity observations indicate 
that hot Jupiters are minor \citep[e.g.,][]{mml11}. Zone 2 represents a distinct deficit of gas giants and it is likely that both the radial velocity and 
transit observations confirm this trend. Most gas giants  pile up in Zone 3. This trend is currently inferred only from the radial velocity observations. 
Zone 4 is for distant gas giants that are detected mainly by the direct imaging method. The most rapidly growing population consists of low-mass planets 
with tight orbits (Zone 5). They are also known as super-Earths as well as hot Neptunes. Recently, these planets have received considerable attention 
because their formation mechanism is unclear and because of their potential importance as abodes for life. 

Several well known theoretical models for explaining the statistical properties of observed exoplanets 
adopt a population synthesis approach \citep{il04i,il08,il10,mab09}.  
The physical picture is based on the core accretion scenario in which the formation of gas giants takes place due to sequential accretion of dust and gas: 
planetary cores form due to runaway and oligarch growth \citep[e.g.,][]{ws89,ki02}, which is followed by gas accretion onto the cores  \citep[e.g.,][]{p96,lhdb09}. 
Population synthesis calculations generate the population of planets by performing a tremendous number of simulations with randomly selected initial conditions. 
The calculations have pioneered an important method for linking theories with observations. Nonetheless, it is still not clear how these
exoplanet populations  arise.

One of the most important problems in theories of planet formation is that planetary migration arises from tidal interactions between protoplanets and 
their natal disks \citep{ward97,ttw02}. The most advanced models of angular momentum exchange between planets 
and homogenous disks show that planetary migration is very rapid ($\sim 10^5$ yr) and that
its direction is highly sensitive to the disk properties such as the surface density and the disk temperature \citep[e.g.,][]{pbck09}. 
This is type I and is distinguished from type II migration that takes place when planets become massive enough to 
open up a gap in their disks. The problem of type I migration is confirmed in population synthesis calculations which show that planetary systems form 
only if the type I migration rate is reduced by some physical process,  by at least a factor of 10 \citep[e.g.,][]{il08}. 

Our recent work, \citet[hereafter, Paper I]{hp12} provides a physical explanation for the concentration of gas giants around 1 AU (Zone 3) and 
the presence of super-Earths and hot Neptunes (Zone 5). The focus of our model is on specific sites in inhomogenous protoplanetary disks where
the net torque vanishes so that  
rapid type I migrators can be captured. The sites, often referred to as planet traps  \citep{mmcf06}, arise from 
the high sensitivity of the torques that drive type I migration, to disk structure \citep{mg04,mpt07,mpt09,hp09b,hp10,lpm10}. 

In Paper I, we considered 
dead zone, ice line, and heat transition traps, and showed that planet traps play two important roles: they capture low mass 
planetary cores and transport them slowly from large to short orbital periods, on the long time scale related to the 
gradual decrease in the disk accretion rate \citep[also see][hereafter, HP11]{hp11}. This combination of planetary growth and transport in traps continues
until planetary mass exceeds the gap-opening mass. At this point in its evolution in the mass-semimajor axis diagram, a planet
undergoes Type II migration with accreting further mass - resulting in radial inward motion.  The planet reaches its final 
point during its disk evolution phase when the disk is finally dispersed by photoevaporation.  
By coupling the core accretion scenario with disk evolution that is regulated by both turbulent viscosity and 
photoevaporation of gas, we can reproduce the observations very well in the sense that the endpoints of planets' evolution tracks computed 
in the mass-semimajor axis diagram are distributed along the radial velocity observations (Paper I). 

In the present paper, we quantitatively evaluate the link between how protoplanets accrete and evolve in specific planet traps, with 
the end-state exoplanetary populations in the mass-period diagram. 
To accomplish this, we perform a parameter study in which we adopt a semi-analytical model developed in Paper I and 
examine planet formation and migration in disks that have a wide range in their physical properties.   We find that the presence of 
discrete exoplanetary populations is readily reproduced by varying just two parameters - the disk accretion rate and lifetime.

Among our many results are the finding that Zone 3 is the preferred end point for gas giants for a wide variety of disk structure.  We also find that 
a large fraction of super-Earths and hot Neptunes are likely to be formed as "failed" cores of gas giants in low mass disks. 
Thus, our results provide important implications for further observations (see Section \ref{disc}). 

The plan of the paper is as follows. In Section \ref{model}, we summarize a semi-analytical model that is used for following evolutionary 
tracks of planets growing at planet traps. First, we have somewhat modified the details of our treatment of 
photoevaporation (the main process for regulating the final stage of disk evolution) 
given in Paper I. This allows greater generality in the prescription of our statistical treatment in which planet formation and evolution at multiple planet traps 
are investigated.  We then define the specific planet formation frequency (SPFF) in Section \ref{resu1} and 
the (integrated) planet formation frequency (PFF) in Section \ref{resu2}. These two functions allow us to quantitatively estimate 
the contributions of planets produced in each type of planet trap to the planetary populations  in each zone in the mass-semimajor axis diagram 
(see Sections \ref{resu1_1} and \ref{resu2_1}). The ensemble of tracks is generated by using distributions of just two control parameters -
the disk accretion rate and the disk depletion time scale.  In Section \ref{resu3}, we present the results of a parameter study in which a number of stellar and 
disk parameters vary, and discuss the nature and extent of planetary populations that are 
predicted by our theory of planet formation in traps. 
In Section \ref{disc}, we compare our results with the observations and discuss implications that can be derived from our findings. 
Our conclusions are given in Section \ref{conc}.

\section{Semi-analytical models} \label{model}

We describe our models that are utilized for generating evolutionary tracks of planets building in planet traps. We briefly summarize 
them and discuss our modifications. We refer the reader to Paper I for the details. 

\subsection{Basic disk models} 

We use steady accretion disk models as the basis for our model. The models are widely used in the literature for 
modeling proroplanetary disks and are characterized by the following accretion rate:
\begin{equation}
 \dot{M}=3 \pi \nu \Sigma_g= 3 \pi \alpha c_s H \Sigma_g,
 \label{mdot}
\end{equation}
where $\Sigma_g$, $\nu=\alpha c_s H$, $c_s$, and $H$ are the surface density, the viscosity, the sound speed, and the 
pressure scale height of gas disks, respectively. We have adopted the famous $\alpha$-prescription for quantifying the 
disk turbulence \citep{ss73}.

\subsection{Properties of planet traps}

\newlength{\myheight}
\setlength{\myheight}{0.5cm}
\newlength{\myheighta}
\setlength{\myheighta}{1cm}
\begin{table*}
\begin{minipage}{17cm}
\caption{Properties of planet traps}
\label{table1}
\begin{tabular}{cccc}
\hline
\parbox[c][\myheight][c]{0cm}{}  Planet trap           &  Position         &  Condition      &  Solid density   \\ 
\hline
\parbox[c][\myheighta][c]{0cm}{}  Dead Zone          & $\frac{r_{dz}}{r_0} = \left( \frac{\dot{M}}{3 \pi (\alpha_A+\alpha_D) \Sigma_{A0} H_0^2 \Omega_0} \right)^{1/(s_A+t+3/2)} $   &  N/A     &  $ \Sigma_{d,dz} \approx \frac{2 \dot{M} f_{dtg}}{3 \pi (\alpha_A+\alpha_D) H^2 \Omega}  $  \\   
\parbox[c][\myheighta][c]{0cm}{} Ice line                 &  $\frac{r_{il,\mbox{H}_2\mbox{O}}}{r_0} = \left[ \frac{1}{T_{m}^{12}(r_{il,\mbox{H}_2\mbox{O}})} \frac{\bar{C}_1 \bar{\kappa}_{il} \Omega_0^3} {\alpha_D}\left( \frac{\dot{M}}{3 \pi} \right)^2 \right]^{2/9} $  &   $ \frac{r_{il,\mbox{H}_2\mbox{O}}}{r_{dz}} >  \left( \frac{H}{r}(r_{dz}) \frac{\alpha_A + \alpha_D}{\alpha_A - \alpha_D} \right)^{1/(s_A+t/2+1)}$  & $ \Sigma_{d,il} \approx \frac{\dot{M} f_{dtg}}{3 \pi \alpha_A H^2 \Omega} $ \\
\parbox[c][\myheighta][c]{0cm}{} Heat transition     &  $ \frac{r_{ht}}{r_0}  = \left[ \frac{1}{T_{m0}} \left( \frac{r_0}{R_*} \right)^{3/7} \left( \frac{\bar{C}_1 \bar{\kappa}_{ht}\Omega_0^3}{\alpha_A} \left( \frac{\dot{M}}{3 \pi} \right)^2  \right)^{1/3}  \right]^{14/15} $ & $r_{ht}>r_{dz}$ &
$ \Sigma_{d,ht} \approx \frac{\dot{M} f_{dtg}}{3 \pi \alpha_A H^2 \Omega}$ \\
\hline
\end{tabular}
$\bar{C}_1 = 1.48 \times 10^{-4}$ in cgs units, the opacity $\bar{\kappa}_{il}= 2\times 10^{16}$ cm$^2$ g$^{-1}$, the condensation temperature of water $T_{m}(r_{il,\mbox{H}_2\mbox{O}})=170$ K, $ T_{m0} \simeq  ( 1/H )^{2/7} ( (\bar{C}_2 T_* R_*)/M_*)^{1/7} T_*$, $\bar{C}_2=5.38 \times 10^{14}$ in cgs units, and the opacity $\bar{\kappa}_{ht}= 2 \times 10^{-4}$ cm$^2$ g$^{-1}$. Also see Table \ref{table3} for the definition of variables.
\end{minipage}
\end{table*}

Our model focuses on the fact that protoplanetary disks should have several types of inhomogeneities
as shown by a number of analytical and numerical studies (see HP11, references 
herein).  Disk inhomogeneities act as trapping sites for protoplanets that undergo rapid type I 
migration through their natal disks \citep{mmcf06}. This is a combined consequence of the distortion of disk properties due to the disk 
inhomogeneities and the high sensitivity of planetary migration to them (e.g., HP11). One kind of disk inhomogeneity is an ice line, 
and its existence in disks is currently being inferred by observations. For example, \citet{qdo11} have recently revealed through Submillimeter Array 
observations that condensation of CO leads to a considerable jump in column density of $^{13}$CO at a CO-ice line. 

We consider three types of disk inhomogeneities and their associated planet traps (see Table \ref{table1}): dead zones, ice lines, and 
heat transitions (HP11, Paper I). Dead zones are present in the inner region of disks where MHD turbulence initiated by magnetorotational 
instabilities (MRIs) is suppressed significantly. The MRI, which arises from the penetration of high energy photons such as X-rays from the central 
stars and cosmic rays,  is substantially diminished in the inner regions of the disk due to the high column density there \citep[e.g.,][]{g96,mp06}. 
As a result, the degree of disk 
ionization becomes low and MRIs are suppressed in subsurface regions within the dead zones. Ice lines are located where certain 
molecules freeze-out onto dust grains in disks due to low disk temperatures relative to the condensation temperature of the molecules 
\citep{sl00,mdk11}. This freeze-out process generates opacity transitions, because the dust is the main absorber of stellar irradiation 
\citep[e.g.,][]{dhkd07}.  In this paper, we take into account a water-ice line.\footnote{CO is the second most abundant molecules in protoplanetary disks. 
Nonetheless, we neglect the CO-ice line, because its position is generally beyond 10 AU where the timescale of oligarchic growth is longer than 
the disk lifetime.} Heat transitions are defined where the main heat source of protoplanetary disks changes from viscous heating to 
stellar irradiation \citep{dccl98,mg04}. This arises because viscous heating is dominant in the inner region of disks and the resultant 
temperature profile is steep whereas stellar irradiation is important for the outer part of disks and ends up with shallower profiles. 
We summarize all the key properties of planet traps in Table \ref{table1} and refer the reader to Paper I for the detail (also see Table \ref{table3}).

\subsection{Time evolution of disks} \label{model_1}

Disk evolution is regulated by at least  two important physical processes \citep[e.g.,][]{a11}. One of them is a diffusive process 
that leads to a slow ($\sim$ Myr) evolution of disks. The other is a dissipative process that causes the depletion of the gas disks, especially in the final stage
of disk 
evolution.  Following Paper I, we take both of these processes into account for modeling time evolution of protoplanetary disks. 

For the diffusive process, we adopt viscous diffusion that is excited by some kind of turbulence in disks.  Specifically, we combine 
similarity solutions of \citet{lbp74} with observational results, which gives the following scaling law for the disk accretion rate (Paper I);
\begin{equation}
 \dot{M}(\tau) \simeq  10^{-8} M_{\odot} \mbox{ yr}^{-1} \eta_{acc} 
                                    \left( \frac{M_*}{0.5M_{\odot}}\right)^2
                                    \left( \frac{\tau}{ \tau_{vis}} \right)^{(-t+1)/(t-1/2)},
 \label{mdot_vis}
\end{equation}
where the typical mass of classical T Tauri stars is assumed to be $\sim 0.5 M_{\odot}$, and the viscous timescale ($\tau_{vis}$) is set as $10^6$ yr, 
and where the disk temperature follows the standard power law form $T\propto r^{t}$  with $t=-1/2$.   
Note that this choice of $t$ reproduces the standard treatment of $\nu \propto r$, since $\nu \propto r ^{(t+3/2)}$ (see Paper I). 
We adopt a fiducial disk accretion rate of $10^{-8} M_{\odot} yr^{-1}$ for these calculations and  introduce a 
dimensionless scale factor for the accretion rate  $\eta_{acc}$.  The distribution of values of this parameter controls
the range of  disk accretion rates which in turn, has observable consequences on the resultant planetary populations. 
  
In order to include dissipation into disk evolution, we modify the treatment of photoevaporation that we used in Paper I.  
Adopting an analytical approximation of the photoevaporation rate derived by \citet{ahl04}, 
we formulated a scaling law for the photoevaporation rate $\dot{M}_{pe}$ (see Equation (23) in Paper I). The photoevaporation 
terminates the time evolution of disks: it was assumed that the disk mass declines with time following viscous evolution (Equation (\ref{mdot_vis})) and 
that the gas disk totally disappears instantly when the accretion rate $\dot{M}$ becomes equal to the photoevaporation rate $\dot{M}_{pe}$. 
As discussed in Paper I, this approximation provides a reasonable estimate for the disk lifetime, compared with more detailed simulations. 

The evolution of disks due to the combination of these two effects results in  the so-called "two-timescale" nature of disk clearing.
Many detailed numerical simulations show that it is a generic feature of disk evolution 
\citep[e.g.,][]{cgs01,acp06b,gdh09,oec11}. Such disk dissipation is consistent partially with the observations of protoplanetary disks which infer 
that the transition from gas to debris disks is likely to occur in $\sim 10^5$ Myr \citep[e.g,][]{wc11}. On the other hand, the recent observations 
also detect "transition" disks that were originally identified from the lack of near-IR excess in spectral energy distributions (SEDs). 
The formation mechanism(s) of transition disks is still unclear, but the disks are currently considered to be in an intermediate phase between 
gas and debris disks \citep[e.g.,][]{znh11,znd12,dhr12}. It is important that some transition disks show good evidence of gas accretion onto the central stars. 
The recent photoevaporation models cannot fully explain such disks, because the two-timescale nature of disk clearing leads to rapid ($\sim 10^5$ yr) 
dissipation of the inner disk after a gap is opened.\footnote{Recently, \citet{m12} have investigated photoevaporation of protoplanetary disks with dead zones 
and shown that photoevaporation may also be able to explain transition disks that have gas accretion onto the central stars, if dead zones are present in 
disks. } Thus, it is unlikely that photoevaporation is only a process of dissipating gaseous protoplanetary disks, 
and hence it is required to improve the treatment of a dissipation process in our model.

\begin{figure}
\begin{center}
\includegraphics[width=9cm]{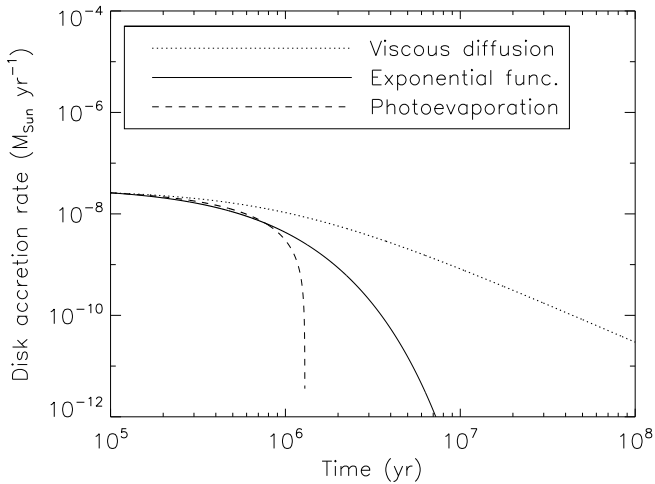}
\caption{Time evolution of the disk accretion rate ($\dot{M}$) governed by viscous diffusion and photoevaporative dissipation of the disk. 
The behavior regulated only by viscous diffusion is denoted by the dotted line 
(see equation (\ref{mdot_exp}) without the exponential function) whereas the behavior regulated by both viscous diffusion and photoevaporation is 
shown by the dashed line (see equation (\ref{mdot_pe2})). Since these two evolutions are likely to provide two extreme cases in the behavior of the disk 
accretion rate, we adopt the rate regulated by equation (\ref{mdot_exp}). Given the uncertainties in disk clearing, equation (\ref{mdot_exp}) represents 
an intermediate case (see the solid line).}
\label{fig2}
\end{center}
\end{figure}

Viscous diffusion and photoevaporation are the two extreme limits for the end stage of disk evolution. 
Figure \ref{fig2} shows how the disk accretion rate evolves with time under the action of each of these processes. When only viscous diffusion 
is taken into account, the accretion rate gradually decreases, following the similarity solutions (see the dotted line). To model this situation, we have adopted 
equation (\ref{mdot_exp}) without the exponential function, assuming that $\eta_{acc}=1$. When photoevaporation acts as a dominant dissipative process, 
the disk accretion rate drops more rapidly (see the dashed line). This is the two-timescale nature of disk clearing. We have adopted equation (\ref{mdot_pe2}) 
in order to model this combined behavior 
(see Appendix \ref {app1} for details). Thus, the former gives the upper limit in the disk accretion rate, especially at the final stage of disk evolution, 
whereas the latter does the lower limit.   

We now proceed with our implementation of a two timescale approach. 
One of the simplest assumptions 
of how the disk mass ($M_{disk}$) dissipates with time is that $d M_{disk}/(dt) = - M_{disk}$, 
following an often used scaling in the population synthesis models developed by \citet{il04i,il08,il10}.
Thus, we adopt the following formula to representing the the time evolution of the disk 
accretion rate in disks that undergo both viscous evolution and (some kinds of) dissipation: 
\begin{eqnarray}
 \label{mdot_exp}
 \dot{M}(\tau) & \simeq & 3 \times 10^{-8} M_{\odot} \mbox{ yr}^{-1} \eta_{acc} 
                                                   \left( \frac{M_*}{0.5M_{\odot}}\right)^2 \\
              &    &  \times \left(1+ \frac{\tau}{ \tau_{vis}} \right)^{(-t+1)/(t-1/2)}  \exp \left( - \frac{\tau-\tau_{int}}{\tau_{dep}}  \right),  \nonumber                                   
\end{eqnarray}
where $\tau_{dep}$ is the depletion timescale of the gas disk and $\tau_{int}=10^5$ yr is the initial time of our computations. 

As shown in Figure \ref{fig2} and as expected,  the disk accretion rate regulated by equation (\ref{mdot_exp}) undergoes an intermediate 
behavior (see the solid line), compared with the limiting cases of viscous diffusion and photoevaporation. Thus, we adopt equation (\ref{mdot_exp}) for 
characterizing the behavior of the disk accretion rate in which the disk evolution over $\sim$ Myr is mainly regulated by turbulent 
viscosity whereas the disk lifetime is likely to be determined by some kind of dissipate processes that are modelled by exponential functions. 

We examine how different treatments of the disk accretion rate affect the resultant planetary populations  in Appendix \ref{app1}.
Specifically, we examine there how sensitive our results are to  the distributions of the characteristic timescales in the problem (introduced
in  Section \ref{resu1_1}) and show that the exact "shape" of the disk accretion rate does not matter. In fact, the distribution of 
the disk lifetime is the most important (summarized in Tables \ref{table4} and \ref{tableA.3}).
We adopt equation (\ref{mdot_exp}) as one of the "representatives" in disk evolution (also see Section \ref{pff_extreme}).

Currently, there is no reliable estimate of when planets start forming within protoplanetary disks (see $\tau_{int}$ in equation (\ref{mdot_exp})). 
We therefore choose a reasonable time based upon the following three considerations. First, we found that 
the results are qualitatively similar if $\tau_{int}$ is smaller than $\tau_{vis}$. Second, we performed a parameter study on $\tau_{vis}$ 
which varies from $10^5$ yr to $10^7$ yr (see Section \ref{resu3_1}). Third, simulations by \citet{jomk07} indicate
that planetesimals out of which Jovian cores are constructed appear within $10^5$ yrs.  Therefore, it is very 
reasonable to adopt the value of $\tau_{int}=10^5$ yr for all the calculations 
done in this paper. We have inserted $\tau_{int}$ in the above equation in order to adjust the value of $\dot{M}$ at the time $\tau_{int}$. 
We have also put a factor of order unity in the bracket that characterizes viscous evolution, because the formula is more accurate for similarity solutions. 
This modification requires us to insert a factor of 3 in the equation for identifying the value of equation (\ref{mdot_exp}) with equation (\ref{mdot_vis}) 
at a time $\tau=10^6$ yr. 

As an aside, we note that the scaling of $\dot{M}$ with $\eta_{acc}$ is essentially identical to varying the total disk mass (see equation (\ref{mdot_exp})). 
Since the disk lifetime is determined largely by the depletion timescale ($\tau_{dep}$), the total disk mass ($M_{disk}$) can be estimated as 
\begin{equation}
M_{disk}= \int d \tau \dot{M}(\tau) \simeq \dot{M}(\tau) \tau_{dep}.
\label{M_tot}
\end{equation}
Thus, for three important quantities characterizing disk properties ($M_{disk}$, $\dot{M}$, and $\tau_{dep}$), any two are independent. 
For reference purposes, Table \ref{table2} summarizes their values. In this paper, we refer to $\eta_{acc}$ as the disk accretion rate parameter 
(rather than the disk mass parameter), because it is more straightforward in our formalism.

\begin{table}
\begin{center}
\caption{Summary of $\dot{M}$, $\tau_{dep}$, and $M_{disk}$}
\label{table2}
\begin{tabular}{ccc}
\hline
$\dot{M}$ ($M_{\odot} yr^{-1}$)   &  $\tau_{dep}$ (yr)  & $M_{disk}$ ($M_{\odot}$)   \\ \hline
$10^{-7}$                                     &  $10^5$               &  0.01                                    \\
$10^{-8}$                                     &  $10^6$               &  0.01                                    \\  \hline
\end{tabular}
\end{center}
\end{table}

\subsection{Planet formation and migration}

We adopt the same model as Paper I for forming gas giants \citep[see Appendix B of Paper I, also see][]{il04i}. The model 
is based on the core accretion picture as the general theoretical framework.  The formation of planetary cores proceeds due to oligarchic growth \citep{ki02}.
In particular, the solid density at planet traps ($\Sigma_{d} \propto r^{-t-3/2}$ with $t=-1/2$), that evolves with time following equation 
(\ref{mdot_exp}), is used for specifying the local growth of planetary cores 
(see Table \ref{table1}). Once the core formation is complete, then the subsequent gas accretion onto the cores takes place that is regulated by the 
Kelvin-Helmholtz timescale \citep{ine00}.

Currently, there is no reliable analytical and numerical study for constraining the final mass 
of planets. In Paper I, we terminated planetary growth artificially when planets obtain 10 percent of the total disk mass at that time. 
In this paper, we improve on this by using a more physically based criterion, namely, that planetary growth stops when planets acquire 
$f_{fin}$ times the gap-opening mass: 
\begin{equation}
 \label{M_max}
 M_{fin} \equiv   f_{fin} M_{gap},
\end{equation}
where the gap opening mass ($M_{gap}$) is given as
\begin{equation} 
          M_{gap} =   \mbox{min}\left[ 3 \left( \frac{H}{r}(r_p) \right)^3, \sqrt{40 \alpha  \left( \frac{H}{r}(r_p) \right)^5} \right] M_*,                              
\end{equation}
where $r_p$ is the position of planets. We set $f_{fin}=10$ for the fiducial case. This choice is motivated by the recent numerical studies 
which show that a considerable amount of gas flows into the gap even after a clear gap is open in the gas disks \citep[e.g.,][]{lsa99,ld06}, which may 
lead to further growth of planets. We will perform a parameter study on it in Section \ref{para_f_fin}.

For planetary migration, we also adopt the same approach as Paper I (see Section 6.2 in Paper I) in which planetary migration occurs in
four distinct stages: slower type I, trapped type I, standard type II, and slower type II migration. When the mass of planets is smaller than the 
minimum mass of planets whose type I migration rate is comparable to the moving rate of planet traps, they undergo slower type I migration, 
and hence the change in their radial position is almost negligible. If planets are more massive than the minimum mass but less massive than 
the gas-opening mass, then trapped type I migration is used for the orbital evolution of the planets. In this case, planets move at the same
rate as their host planet traps. When planets grow to be more massive than the gap-opening mass but less than the local disk mass 
defined by $M_{crit}=\pi \Sigma_g(r_p) r_p^2$, the planets migrate inwards due to the standard type II, wherein the migration timescale is 
characterized by the local viscous timescale of gas disks. Once planets acquire masses more than $M_{crit}$, 
then the inertia of the planets becomes effective and standard type II migration slows down by a factor of $\sim M_{crit}/M_p$. 
The orbital evolution of planets is stopped artificially if the position of the planets reaches $r=0.04$ AU, following \citet{il04i,il08}.

We terminate our calculations either when the time $\tau$ reaches $10^9$ yr or when the disk accretion rate ($\dot{M}$) declines to 
$10^{-14} M_{\odot}$ yr$^{-1}$. We confirmed that our results are not affected by these two conditions.

\section{Specific planet formation frequencies} \label{resu1}

Based the semi-analytical model discussed above, we quantify how planet traps can generate specific planetary populations in the mass-period diagram. 
In order to proceed, we define specific planet formation frequencies (SPFFs), as discussed below.

\subsection{Parameters}

\begin{table*}
\begin{minipage}{17cm}
\begin{center}
\caption{Important quantities}
\label{table3}
\begin{tabular}{ccc}
\hline
Symbols            &  Meaning                                                                                                                                     & Fiducial values           \\ \hline
                          &  Model parameters                                                                                                                     &                                    \\ \hline
$\Sigma_{A0}$  &  Surface density of active regions at $r=r_0$                                                                             & 20  g cm$^{-2}$          \\           
$s_A$                &  Power-law index of $\Sigma_A (\propto r^{s_A})$                                                                    & 3                                  \\
$\alpha_{A}$      &  Strength of turbulence in the active zone                                                                                 & $10^{-3}$                    \\        
$\alpha_{D}$     &  Strength of turbulence in the dead zone                                                                                   & $10^{-4}$                     \\  
$t$                     &  Power-law index of the disk temperature ($T \propto r^{t}$)                                                     & -1/2                               \\  
$f_{fin}$            &  Final mass of planets (see Equation (\ref{M_max}))                                                          & 10                                  \\ \hline
                         &  Stellar parameters                                                                                                                      &                                       \\ \hline
$M_*$               &  Stellar mass                                                                                                                                & 1 $M_{\odot}$               \\
$R_*$               &  Stellar radius                                                                                                                                &  1 $R_{\odot}$              \\
$T_*$               &  Stellar effective temperature                                                                                                        & 5780 K                           \\ \hline
                         &  Disk mass parameters                                                                                                                 &                                       \\ \hline
$\eta_{acc}$     &  a dimensionless factor for $\dot{M}$ (see Equation (\ref{mdot_exp}))                                         &                                    \\
$f_{dtg}$           &   the dust-to-gas ratio ($=f_{dtg, \odot} \eta_{dtg} \eta_{ice}$)                                                     &                                    \\
$f_{dtg, \odot}$ &   the dust-to-gas ratio at the solar metallicity                                                                                & 6 $\times10^{-3}$      \\
$\eta_{dtg}$      &   A parameter for increasing/decreasing  $f_{dtg}$                                                                      &  1                                  \\
$\eta_{ice}$      &   A parameter for increasing/decreasing $f_{dtg}$ due to the presence of ice lines                     &                                      \\ \hline
                         &  Disk lifetime parameters                                                                                                               &                                      \\ \hline
$\tau_{vis}$      &  the viscous timescale (see Equation (\ref{mdot_exp}))                                                                 &   $10^6$ yr                  \\                         
$\tau_{dep}$      &  the depletion timescale ($=\tau_{dep,0} \eta_{dep} $, see Equation (\ref{mdot_exp}))      &                                      \\
$\tau_{dep,0}$  &  the typical depletion timescale for the fiducial case                                                         &   $10^6$ yr                    \\
$\eta_{dep}$       &   A parameter for increasing/decreasing  $\tau_{dep}$                                                                   &                                        \\
\hline
\end{tabular}
\end{center}
\end{minipage}
\end{table*}

Fundamental parameters for regulating planet formation in protoplanetary disks can be categorized into three sets: stellar parameters, 
disk mass parameters, and disk lifetime parameters. Population synthesis calculations confirm the importance of these three sets of parameters 
for understanding the observations of exoplanets \citep{il04i,il08,mab09,amb11}. In addition to them, we consider one additional class. 
Table \ref{table3} summarizes these parameters. 

Among the disk mass parameters, we assume that the dust-to-gas ratio can be scaled as 
\begin{equation}
f_{dtg}=f_{dtg,\odot} \eta_{dtg} \eta_{ice},
\end{equation}
where $f_{dtg,\odot}$ is the the dust-to-gas ratio for solar metallicity and $\eta_{ice}$ is a factor for increasing/decreasing the dust-to-gas ratio 
due to the presence of ice lines. When planet traps are within an ice line, we set $\eta_{ice}=1$, because there is no increase (or decrease) in $f_{dtg}$ 
due to the ice line. When traps are on and beyond the ice line,  one expects that $f_{dtg}$ becomes larger due to condensation of a certain 
molecule onto dust grains that arises from the low disk temperature \citep[e.g.,][]{sl00,mdk11}. We adopt the widely used values in the literature for 
quantifying $\eta_{ice}$: $\eta_{ice}=3$ when traps are beyond the ice line, and $\eta_{ice}=4$ for the ice line trap \citep{h81,phbs94,il04i}. 
Since the dust-to-gas ratio is currently most likely to be considered as $\sim 1.8 \times 10^{-2} $ at the solar metallicity \citep{ags09}, this ends up 
as $f_{dtg,\odot} = 6 \times 10^{-3}$. 

We also parameterize the disk depletion timescale $\tau_{dep}$ by introducing a dimensionless parameter;
\begin{equation}
 \label{tau_dep}
 \tau_{dep}= \tau_{dep,0} \eta_{dep},
\end{equation} 
where it is assumed that a fiducial depletion timescale is of the order $\tau_{dep,0}=10^6$ yr. 

\subsection{Definition of SPFFs}

As shown in Table \ref{table3}, there are many parameters in these kinds of calculations, which is why the standard population 
synthesis calculations perform a tremendous number ($\ga 10^4$) of simulations with the initial conditions being randomly selected.  
We instead adopt another approach to evaluate the "efficiency" with which planets formed in some trap end up populating some given zone in the
mass-period diagram.  Rather than doing Monte Carlo simulations for the very large range of parameters,  we focus on just two that are particularly 
important - the disk accretion rate and lifetime parameters.\footnote{As pointed out in Section \ref{model_1}, the parameter $\eta_{acc}$ can be viewed 
as the disk mass parameter (see equation (\ref{M_tot}), also see Table \ref{table2}). In fact, we categorize $\eta_{acc}$ into the disk mass parameter in 
Table \ref{table3}. Nonetheless, it is straightforward to call $\eta_{acc}$ the disk accretion rate parameter in our formalism 
(see equation (\ref{mdot_exp})). Therefore, $\eta_{acc}$ is referred to as the disk accretion rate parameter in the following sections.}  
We compute evolutionary tracks of planets forming in planet traps in the mass-semimajor axis diagram, as done in Paper I,  controlled by a range of values 
for just these two key parameters. In the most of calculations, we consider the ranges of $\eta_{acc}$ ($0.1 < \eta_{acc} < 10$) and $\eta_{dep}$ 
($0.1 < \eta_{dep} < 10$). Since we control how many tracks ($N$) are followed in our calculations, we can define the SPFF as below;
\begin{equation}
\label{spfr}
\mbox{SPFFs(Zone i, } \eta_{acc}, \eta_{dep}) \equiv \frac{N\mbox{(Zone i, } \eta_{acc}, \eta_{dep})}{N_{int}},
\end{equation}
where $N \mbox{(Zone i, } \eta_{acc}, \eta_{dep})$ is the number of tracks that end up in Zone i after the calculations are done and 
$N_{int}$ is the total number of tracks that are considered in the calculations. We call this a "{\it specific} planet formation frequency", 
because the value is derived from particular {\it given} values for the disk accretion rate ($\eta_{acc}$) and lifetime ($\eta_{dep}$) parameters. 

\subsection{Initial conditions}

In order to estimate the SPFF, we adopt the initial conditions that are similar to those of Paper I (see Section 6.4 in Paper I). In particular, 
we assume that planetary cores start growing from  $\simeq$ 0.01 $M_{\oplus}$ at a position $r$ and a time $\tau$. We confirmed that the initial mass 
is small enough, so that our choice of the value does not affect the results. For the initial position (or time), we consider 100 positions (or times) for each 
planet trap. (This means that $N_{int}=300$). We have confirmed that the number is large enough for the results to converge. We note that, although 
it is assumed that the initial positions of cores exactly correspond to those of one of planet traps, this does not assume that the cores are always trapped 
by them initially. 

Although planet formation could be initiated anywhere in the disk, only protoplanets that experience the slowing down of type I migration resulting from 
capture in a trap  can grow to gas giants and contribute to planetary populations in the mass-semimajor axis diagram \citep{il08}.  Thus, it may not be 
necessary to consider planets that plunge into the central star due to rapid type I migration. Our calculations are therefore sufficient for estimating the SPFFs 
and the PFFs (see equation (\ref{pfr}) for the definition).

\section{Results: Specific Planet Formation Frequencies} \label{resu1_1}

We present the results of the SPFFs for a wide variety of disk accretion rate and lifetime parameters, adopting the values given in Tables \ref{table3}. 
Figure \ref{fig3} shows the resultant SPFFs as a function of both the disk accretion rate ($\eta_{acc}$) and the disk lifetime ($\eta_{dep} $) parameters. 
For reference purposes, we also label physical quantities that are involved with $\eta_{acc}$ and $\eta_{dep}$. We estimate the 
corresponding $\dot{M}$, adopting the fiducial values $M_*=1M_{\odot}$, $\tau=10^6$ yr, and $\tau_{dep}=10^6$ yr, as already 
noted. For $\eta_{dep}$, the disk lifetime is relevant and 
the timescale is essentially identical to the depletion timescale in our model ($\tau_{dep}$, see equation (\ref{tau_dep})). 
The SPFFs of Zone 1 are shown in the top left, Zone 2 in the top right, Zone 3 in the bottom left, and Zone 5 
in the bottom right. No planet ends up in Zone 4 in most of the calculations done in this paper. The values of the SPFFs are denoted by 
both contours and colors (see the color bar for the corresponding values of the SPFFs). We plot only the values of the SPFFs that are larger than 0.1, 
because the predominant behaviors of the SPFFs in terms of $\eta_{acc}$ and $\eta_{dep}$ appear in such high values. Also, we find that it is useful 
to apply a kind of "filter" to cut out low values for SPFFs. Note that even when the filter is used, some noise is seen (black spots in 
the two bottom panels - Zones 3 and 5).   We checked that this noise arises from the high sensitivity of planet formation, 
especially to very short disk lifetimes, but that the values are reasonably low and our results converge.

\subsection{Gas giants within 10 AU}

We first discuss the SPFFs of gas giants (Zones 1, 2 and 3). Figure \ref{fig3} shows that the resultant SPFFs of 3 zones are well separated from one
another in terms of the disk lifetime (see the two top and left bottom panels). The SPFFs of Zone 1 preferentially cover the region of long disk lifetimes, and 
those of Zone 2 follow Zone 1 and fill out the region of slightly shorter disk lifetimes. The SPFFs of Zone 3 cover the largest region along the disk lifetime axis, 
centered around $\eta_{dep}=1$. 

This behavior is readily understood by two kinds of migration: trapped type I and subsequent type II migration. In our formalism, 
planetary cores form at planet traps and the traps move inward with the cores, following the disk evolution (HP11, Paper I). 
As a result, the distribution of cores of gas giants is achieved in which most cores orbit around 1-10 AU in disks. It is important that 
this "initial" distribution of protoplanets that is generated by planet traps leads to the concentration of gas giants beyond 1 AU. 
When planets form in long-lived disks, type II migration comes into play, and the initial distribution of protoplanets "diffuses" toward the central stars. 
This is because long-lived disks allow planets to have a long time over which they undergo type II migration. The planets therefore tend to end up 
in the vicinity of stars (Zones 1 and 2). Thus, the SPFFs of different zones occupy different regions of the disk lifetime.

Figure \ref{fig3} also shows the vertical structure of the SPFFs of Zones 1, 2, and 3 (see the two top and left bottom panels): all the panels indicate that 
the SPFFs take higher values for higher disk accretion rates, and their values become less than 0.1 for lower disk accretion rates. For example, 
the SPFFs of Zone 1 become lower than 0.1 at $\eta_{acc} \simeq 0.5$, those of Zone 2 at $\eta_{acc} \simeq 0.3$, and those of Zone 3 at 
$\eta_{acc} \simeq 0.2$. Since the disk accretion rate is linearly proportional to the total disk mass in our model (see equation (\ref{M_tot}), also see 
Table \ref{table2}), this can be explained by the core accretion scenario. In the scenario, the formation efficiency of gas giants is obviously high for disks 
with large masses (high accretion rates), because solid materials that form cores of gas giants become abundant relative to the low-mass disks. 
Note that the value of the disk metallicity is fixed ($\eta_{dtg}=1$) in this paper (see Table \ref{table3}). It is interesting that the threshold value of 
$\eta_{acc}$, above which the SPFFs become larger than 0.1, increases from Zone 3, to Zone 2, and up to Zone 1.  

Overall, the results show that the SPFFs of Zone 3 cover the largest parameter region in the disk accretion rate-lifetime diagram, and 
hence suggest that Zone 3 is likely to be the most preferred place for gas giants to end up. This is one of the most important findings of this study and 
confirms one of the conclusions given in Paper I that planet traps play the crucial role for understanding the statistics of observed exoplanets around 
a solar-type of stars. It is interesting that the peak of the SPFFs of Zone 3 is obtained approximately along the fiducial value of the disk lifetime 
($\tau_{dep}=10^6$ yr with $\eta_{dep}=1$). 

\subsection{Distant gas giants}

We find that the SPFFs of Zone 4 are zero for a wide range of the disk mass and lifetime parameters, (so that we do not show the plot in Figure 
\ref{fig3}). Note, however, that our model focuses on disks around T Tauri stars, rather than higher mass Herbige Ae/Be stars, which are the host stars 
for the distant planets (also see Section \ref{resu3_2}). We will, in future, apply our model to disks around Herbig Ae/Be stars.

\subsection{SuperEarths and hot Neptunes}

\begin{figure*}
\begin{minipage}{17cm}
\begin{center}
\includegraphics[width=8cm]{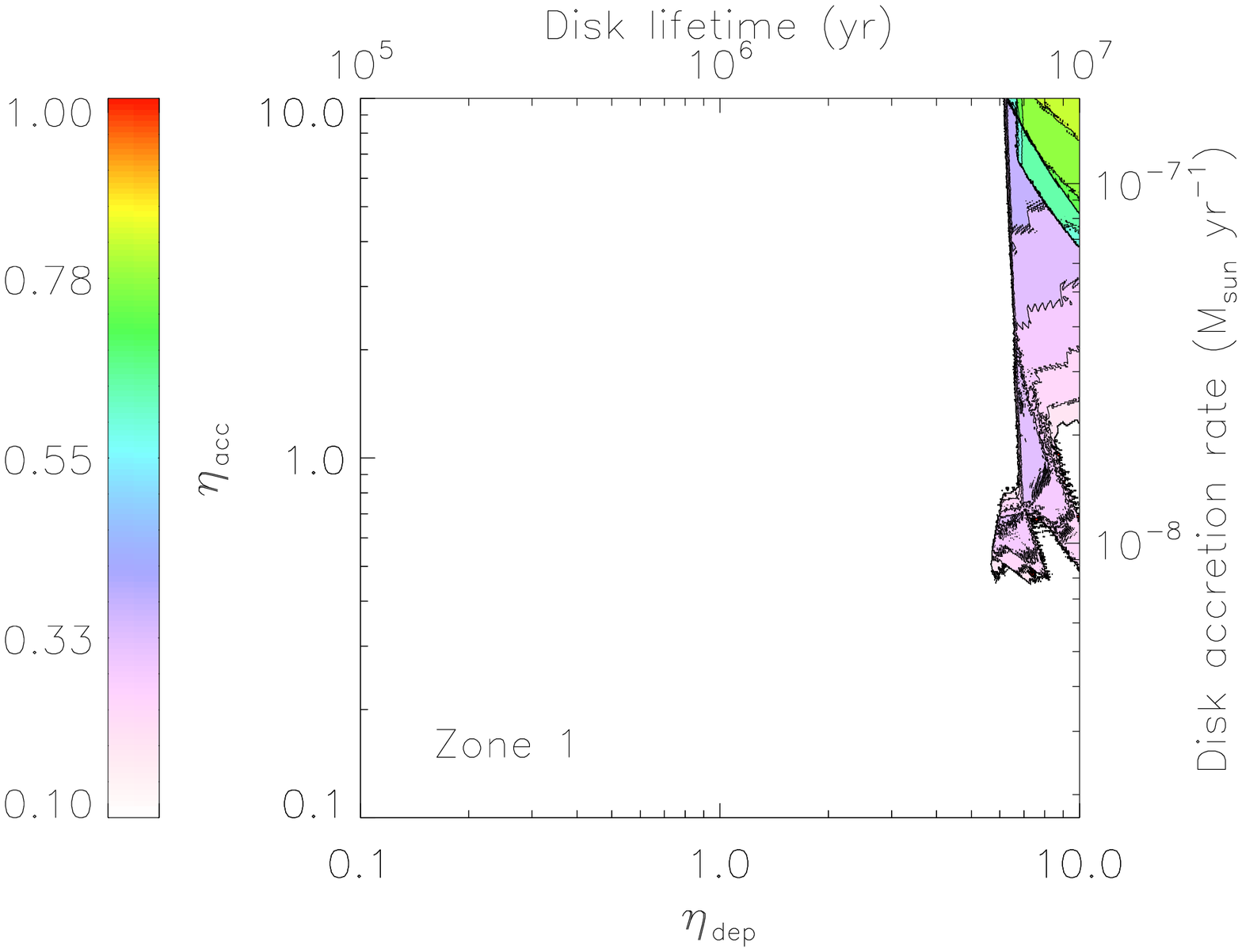}
\includegraphics[width=8cm]{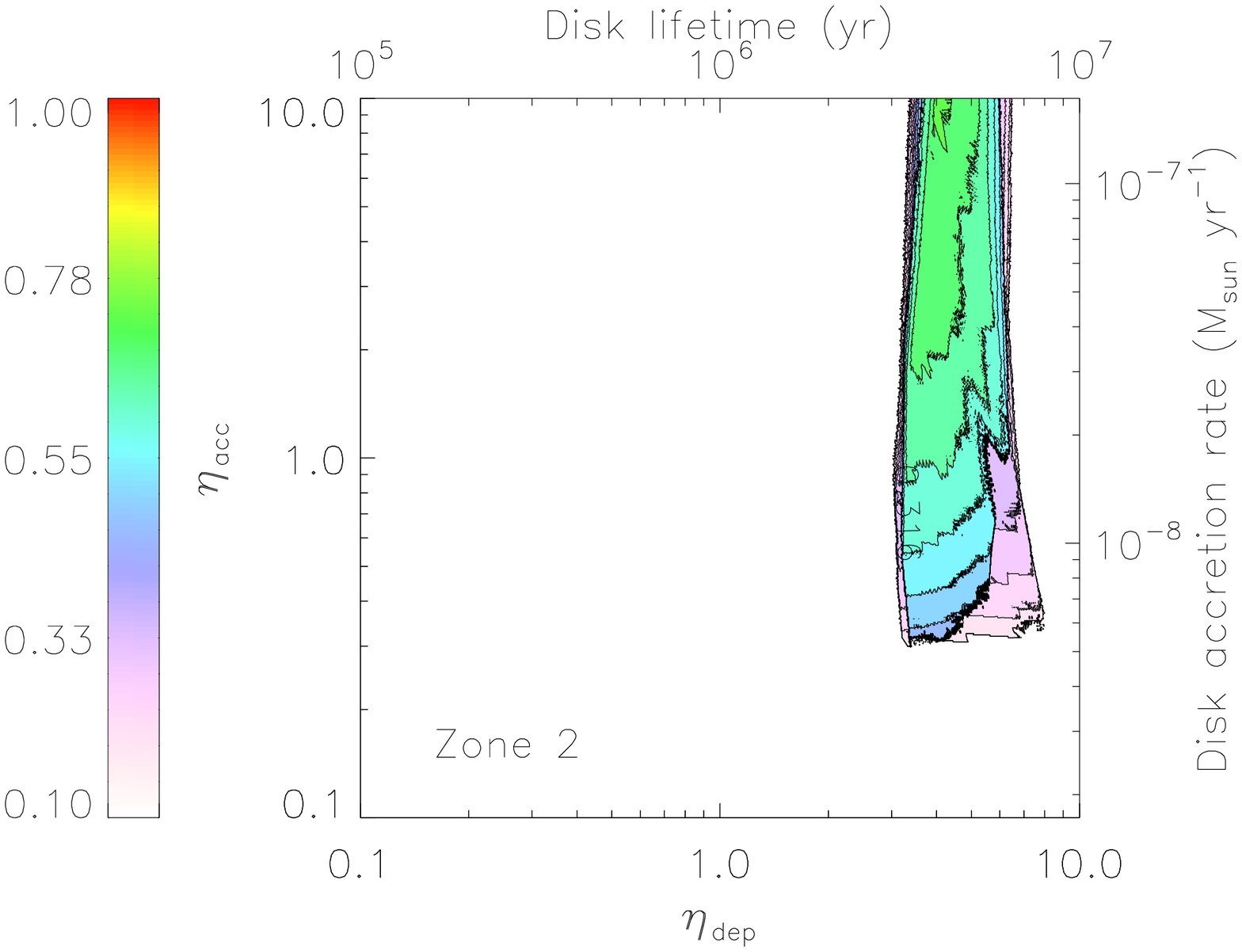}
\includegraphics[width=8cm]{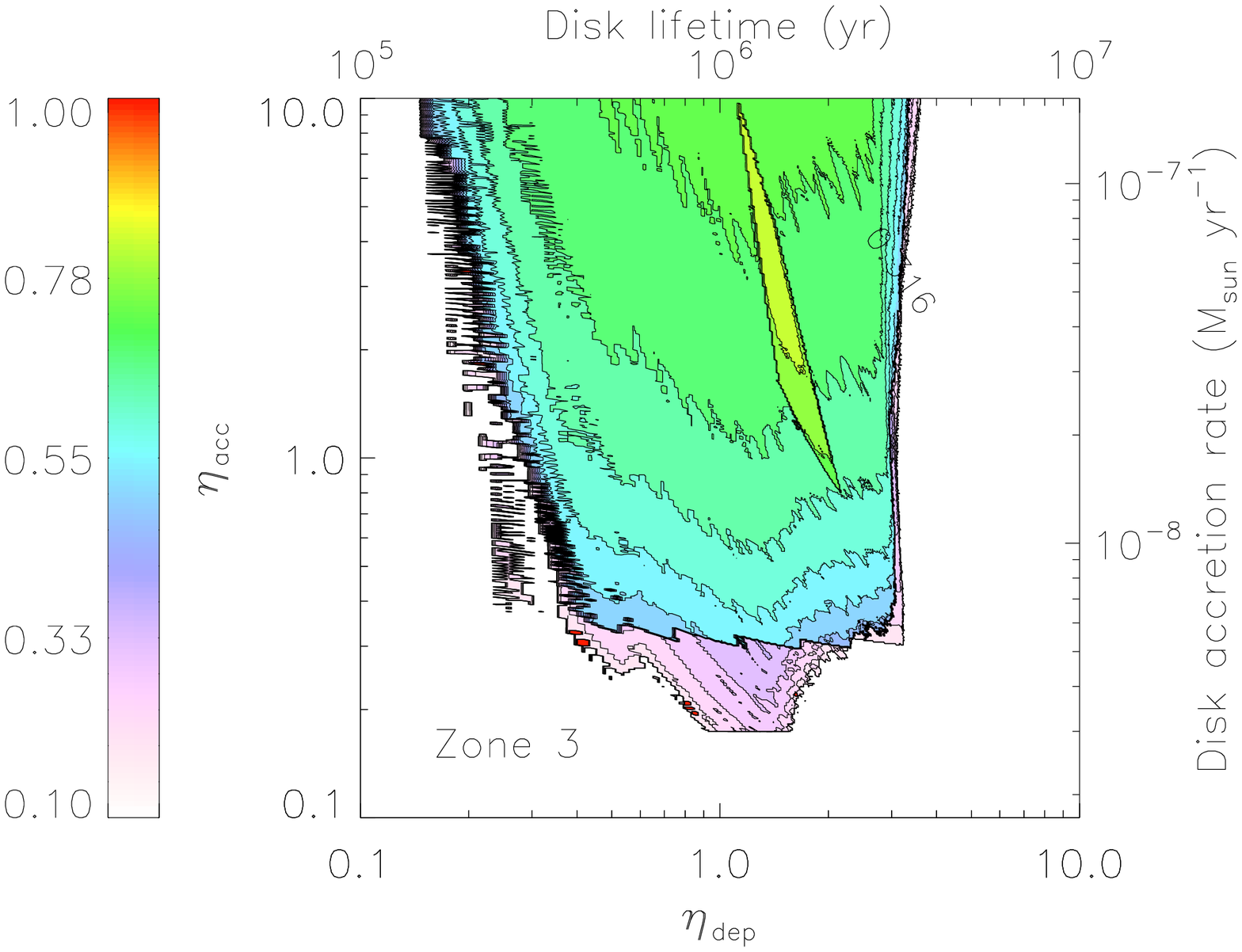}
\includegraphics[width=8cm]{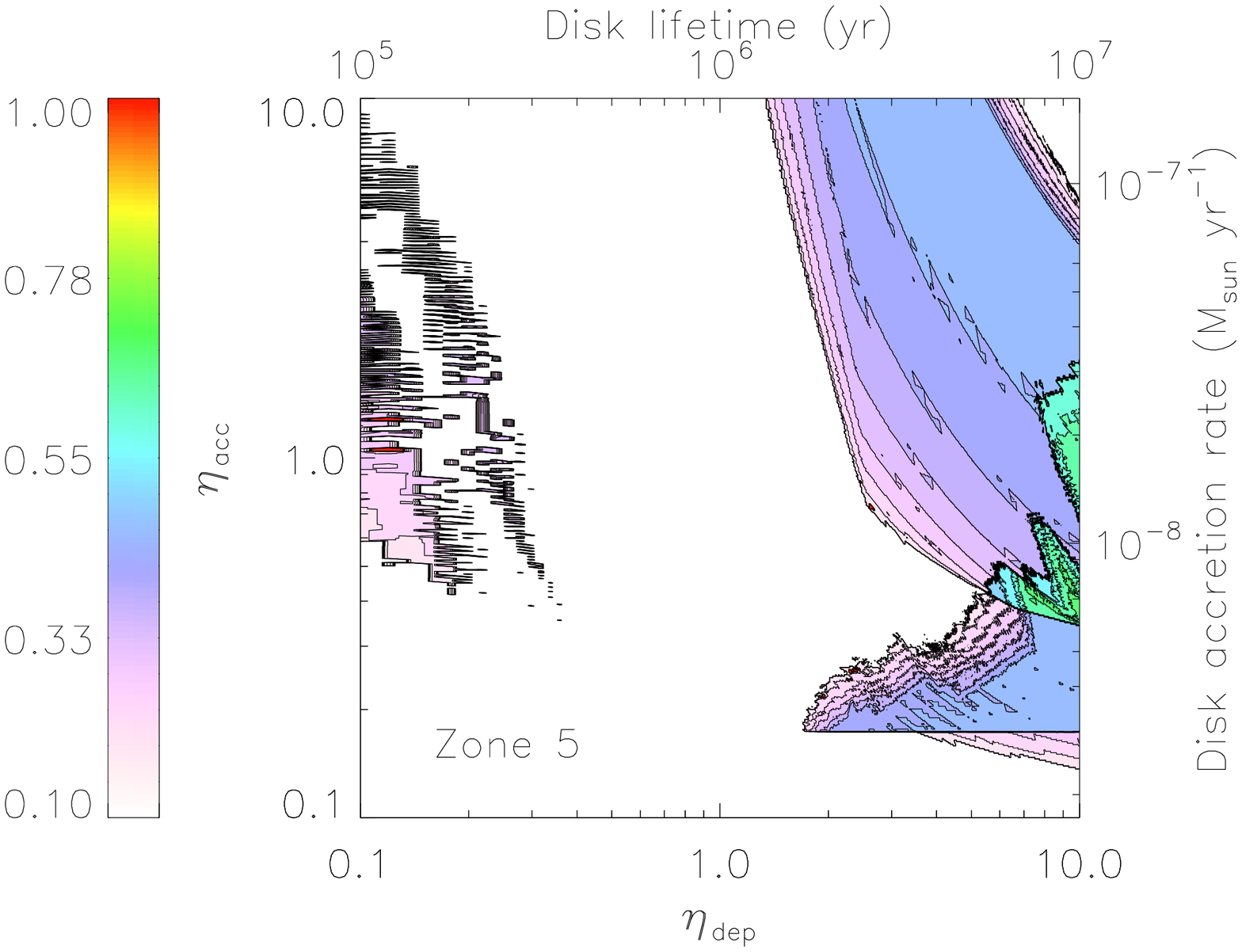}
\caption{The SPFFs as a function of disk accretion rate ($\eta_{acc}$) and depletion time ($\eta_{dep}$). The SPFFs are defined by equation (\ref{spfr}), 
and denoted by the contours and colors (see the color bar for the corresponding value of the SPFFs). For the purpose of the clear presentation, 
we only show the values of the SPFFs that are larger than 0.1. Also, we denote physical quantities that correspond to $\eta_{acc}$ and $\eta_{dep}$ 
as a reference.  The fiducial value of  $\eta_{acc} = 1$,  corresponds to a disk accretion rate of $\sim 1.7 \times 10^{-8} M_{\odot} yr^{-1}$.
The fiducial value of $\eta_{dep}$ corresponds to a disk depletion time of $\tau_{dep}=10^6$ yr.
We have adopted standard values of  $M_*=1 M_{\odot}$, $\tau=10^6$ yr, (see equation (\ref{mdot_exp})).
For $\eta_{dep}$, the disk lifetime is shown, and the timescale is calculated by the depletion timescale ($\tau_{dep}$) in our model 
(see equation (\ref{tau_dep})). The top left panel denotes the SPFFs of Zone 1, the top right is for Zone 2, the bottom left for Zone 3, 
and the bottom right for Zone 5. The SPFFs of gas giants (Zones 1, 2, and 3) essentially do not overlap with each other.}
\label{fig3}
\end{center}
\end{minipage}
\end{figure*}

We discuss the SPFFs of low mass planets in tight orbits (Zone 5). Compared with the cases of gas giants, Figure \ref{fig3} shows 
a more complicated behavior (see the right bottom panel). In fact, the panel shows that the high values of SPFFs appear in 3 distinct areas 
in the disk accretion rate-lifetime diagram: long lifetimes and large accretion rates (Area 1), long lifetimes and low accretion rates (Area 2), 
and short disk lifetimes and low accretion rates (Area 3). We note that these areas are defined in the disk accretion rate-lifetime diagram and 
are totally different from 5 zones defined in the mass-semimajor axis diagram (see Figure \ref{fig1}). Planets experience significant radial drifts 
due to the movement of planet traps and/or subsequent type II migration. As a result, they end up in the vicinity of the host stars (Zone 5). 
Our results show that in order to have low mass planets, the host disks must be on the lower end of the disk accretion rate (disk mass) 
in our parameter space.  

We now examine 3 areas in detail. In Area 1, it is obvious that gas giants form predominantly in most stages of disk evolution and 
they eventually fill out Zone 1. In addition, it is general to expect that disks in Area 1 can also form low-mass planets after the gas giant formation 
is complete. Thus, low-mass planet formation takes place when the accretion rate (mass) of the disks declines. Since disks in Area 1 have 
long disk lifetimes, low-mass planets formed at the later stage of disk evolution can still have enough time to be transported to the proximity 
of the host stars. A similar argument can be applied for Area 2. The only difference with Area 1 is that disks in Area 2 cannot form gas giants 
due to the low accretion rate of disks. This therefore indicates that disks in Area 2 maintain the formation of low-mass planets over the entire disk lifetime. 
Also, their long lifetimes support radial drifts of planets for a long time. One may consider that disks in Area 3 may have difficulties in applying 
the same argument. This is because the disks have short lifetimes. The concern is partially valid in a sense that short disk lifetimes decrease 
the planet formation efficiency. Nonetheless, we can still apply a similar argument, because the rapid decline in the disk accretion rate leads to 
the fast movement of planet traps (see Table \ref{table3}). Thus, planets formed in Area 3 can be delivered to the proximity of stars largely 
by the rapid movement of planet traps. 

The formation of low-mass planets with tight orbits is currently of great interest, because it enables us to discuss the formation mechanism of 
super-Earths and hot Neptunes. It is presently unclear how they form.  Our calculations shed some light on this issue since all the planets are formed via 
the core accretion scenario. Thus, our results imply that many of observed super-Earths and hot Neptunes may be formed as "failed" cores of gas giants 
that cannot accrete gas surrounding the cores. 

\section{Planet formation frequencies} \label{resu2}

\subsection{Definition}

The SPFFs are useful for quantitatively discussing the resultant population of planets formed in planet traps for a wide range of disk accretion rate and 
lifetime paramaters. As we have seen, this two parameter space enables us to examine what kind of disks are preferred for filling up the 5 zones.  
In order to compute the total population of planets that ends up in any zone, we must integrate over these two parameters to find the "(integrated) 
planet formation frequencies (PFFs)" , defined as:
\begin{eqnarray}
 \label{pfr}
 \mbox{PFFs(Zone i)}   & \equiv  &  \\
            \sum_{\eta_{acc}} \sum_{\eta_{dep}}  w_{mass}(\eta_{acc}) &w_{lifetime}(\eta_{dep}) & \mbox{SPFFs(Zone i, } \eta_{acc}, \eta_{dep}),  \nonumber                                   
\end{eqnarray}
where $ w_{\eta_{acc}}$ and $w_{\eta_{dep}}$ are both weight functions of disk accretion rate and lifetime, respectively. The reason why we refer to the value 
as the ({\it integrated}) PFF is that the PFFs are calculated by {\it integrating} the SPFFs over the complete
distributions of disk accretion rate ($\eta_{acc}$) and lifetime ($\eta_{dep}$). 

In principle, the weight functions represent the probabilities of protoplanetary disks with certain accretion rates or lifetimes.
As a result, we can consider the PFFs as the planet formation "efficiency" that is properly weighted by the observations of protoplanetary disks. 
Thus, the PFFs enable us to discuss the statistical properties of planets formed in planet traps without computing population synthesis calculations.

\subsection{Weight functions} \label{weight_func}

We adopt Gaussion functions for both $ w_{mass}(\eta_{acc})$ and $w_{lifetime}(\eta_{dep})$:
\begin{equation}
\label{w_acc}
w_{mass}(\eta_{acc}) = w_{mass,0} \exp \left[ - \frac{1}{2} \left( \frac{\ln \eta_{acc}}{\sigma_{mass}} \right)^2 \right],
\end{equation}
where $w_{mass,0}$ is a normalization constant, $\sigma_{mass}=1$ is the standard deviation, and 
\begin{equation}
\label{w_pe}
w_{lifetime}(\eta_{dep}) = w_{lifetime,0} \exp \left[ - \frac{1}{2} \left( \frac{\eta_{dep} - \mu_{lifetime}}{\sigma_{lifetime}} \right)^2 \right],
\end{equation}
where $w_{mass,0}$ is a normalization constant, $\sigma_{lifetime}=3$ is the standard deviation, and $\mu_{lifetime}=1.5$ is the mean value.

When we select the values of the standard deviation ($\sigma$) and the mean ($\mu$) in equations (\ref{w_acc}) and (\ref{w_pe}), 
we combine the observations of protoplanetary disks and our preliminary results. For $w_{mass}(\eta_{acc})$, we assume that the peak value is achieved 
at $\eta_{acc}=1$ (see equation (\ref{w_acc})). More physically, we pick the values of $\sigma$ and $\mu$, so that the resultant $w_{mass}(\eta_{acc})$ 
can well reproduce the observations which show that the median value of the disk accretion rate is $\sim 10^{-8} M_{\odot}$ yr$^{-1}$ for disks around 
classical T Tauri stars ($M_*\simeq 0.5 M_{\odot}$) of age $\tau \simeq 10^6$ yr \citep[e.g.,][]{hcg98,wc11}. We confirmed that the results are 
quantitatively similar even if we change the value of $\sigma_{mass} (\ge 1)$ with both $\sigma_{lifetime}$ and $\mu_{lifetime}$ fixed. 
For $w_{lifetime}(\eta_{dep})$, we again rely on the observations which statistically show that the disk fraction of stars decreases with time \citep[e.g.,][]{wc11}. 
It is shown that this behavior can be fitted by the exponential function with the e-folding timescale 2.5 Myr \citep{m09}. Since $w_{lifetime}(\eta_{dep})$ 
represents the existence probability of disks, we integrate the equation in terms of $\eta_{dep}$ and find that the fitting can be reproduced reasonably 
when we choose that $\mu_{lifetime}=1.5$ and $\sigma_{lifetime}=3$. We also confirmed that the results do not change very much for a wide value of 
$\sigma_{lifetime}(>3)$, with both $\sigma_{mass}$ and $\mu_{lifetime}$ fixed. Finally, we tried uniform distributions in log scale for both 
$ w_{mass}(\eta_{acc})$ and $w_{lifetime}(\eta_{dep})$, and confirmed that, for the given range of $\eta_{acc}$ $(0.1 \le \eta_{acc} \le 10$) and 
$\eta_{dep}$ $(0.1 \le \eta_{dep} \le 10)$, the results are qualitatively similar to those generated by Gaussian functions with the above values adopted for 
$\sigma$ and $\mu$. Since the Gaussian functions are more likely to be consistent with the observations, we adopt them, rather than the uniform distributions.

\section{Results: Planet Formation Frequencies} \label{resu2_1}

We present the results of the PFFs that are calculated for the fiducial case. Table \ref{table4} summarizes the PFFs for 5 zones. We note that 
the total PFFs are not 100 \%. This is because, when the PFFs are computed, we count only planets that finally distribute into 5 zones. 
The remainder generally ends up below Zones 3, 4, and 5 (see Figure \ref{fig1}). This suggests that a large number of low-mass planets are formed 
over a wide range of semimajor axis in our model. However, we do not discuss these planets for the following two reasons. First, the orbital distribution of 
these planets overlaps with more massive planets that are also formed at planet traps. It is therefore reasonable to anticipate that the results of PFFs for 
these low mass planets would be affected largely by the interactions with the massive planets, which is not modeled in our calculations. Second, 
the mass of such planets is very low, and hence it can be considered that they are embryos of rocky planets. For this case, the final mass of the planets 
can be very different from those estimated from our results, due to the subsequent growth that mainly takes place in planetesimal disks. 
This is also beyond the scope of this paper. 

\subsection{PFFs summed up for three planet traps}

We find that the PFFs in Zone 3 are the highest. This is the most important finding in this paper and is explained by the combination of the SPFFs 
and the weight function ($w_{lifetime}(\eta_{dep})$). As discussed in Section \ref{resu1_1}, the SPFFs of Zone 3 fill out the largest space in 
the disk accretion rate-lifetime diagram, compared with any other zones (see Figure \ref{fig3}). Also, the peak value of the SPFFs corresponds to 
the highest probability of disks (see equation (\ref{w_pe})). As a result, the SPFFs of Zone 3 and $w_{lifetime}(\eta_{dep})$ overlap with each other 
most efficiently.  Thus,  Zone 3 is the preferred place for gas giants to end up when planet formation takes place in disks with planet traps.  

The PFFs of other zones can also be understood by a similar argument. For Zone 1, the SPFFs achieve the maximum value when the disk lifetime 
($\eta_{dep}$) is very long. This minimizes the overlap between the SPFFs and the weight function, and leads  to the lowest value of the PFFs 
(except for Zone 4). For Zone 2, the PFFs are higher than those of Zone 1. This is because the SPFFs are high for disk lifetimes that are slightly shorter 
than the case of Zone 1. This results in the enhancement of the overlap between the SPFFs and the weight function, compared with Zone 1. 

It is interesting that the PFFs of Zone 5 are the second highest for the fiducial case. This is also a consequence of the efficient overlap between 
the SPFFs and the weight function. As shown in Figure \ref{fig3}, the larger area is filled by non-zero SPFFs. Since the SPFFs of Zone 5 are lower 
than 0.1 around $\eta_{dep}=1$, the PFFs of Zone 5 are lower than those of Zone 3. This finding suggests that a large fraction of observed super-Earths 
and hot Neptunes can be regarded as "failed" cores of gas giants.   

\subsection{PPFs for extreme cases: pure viscosity or sharp photoevaporative truncation} \label{pff_extreme}

As documented in the Appendix \ref{app1}, we have run a series of calculations for two extreme cases:  pure viscous evolution (Case 1) and sharply
truncated photoevaporation together with viscosity (Case 2).  The results are shown in Tables \ref{tableA.1} and \ref{tableA.2} for the fiducial case.  
We find that both of these extremes fail to reproduce the observations in interesting ways.
Case 1 extends the disk lifetimes and hence, enhances the PFFs of Zones 1 and 5 significantly (Table \ref{tableA.1}).  On the other hand, 
Case 2 sharply truncates the disk evolution in time, which leaves planets stranded at larger disk radii.  
Thus, we see that planets predominately populate Zone 3.  

\subsection {Sensitivity to distribution of disk lifetimes}

How sensitive are these results to the exact shape of the disk accretion rate, particularly as we vary the distribution of disk lifetimes?  
For example, recent observations of disks imply that the median value of disk lifetimes is about $3$ Myrs with a substantial tail to older ages 
\citep[e.g.,][]{wc11}, which is somewhat longer than our fiducial case.   Does this matter for the resulting populations?
To address this point we focus on long-lived disks and examine whether or not the resultant planetary population generated by Case 2 can 
reproduce the observations of exoplanets.\footnote{We investigate only Case 2. This is because if this would occur for Case 1, then optically thick disks 
would be observed at the age of $>100$ Myrs, which is not the case.} The results are summarized in Table \ref{tableA.3}. Although there are some quantitative 
differences in PFFs between the fiducial and this case, they are qualitatively similar: most formed gas giants end up in Zone 3 whereas low-mass planets in 
tight orbits (Zone 5) are also filled out by planets that fail to accrete gas. 

Thus, we can conclude that the exact "shape" of the disk accretion rate is not important and that the distribution of the disk lifetimes is 
the most crucial parameter for understanding planetary populations in the 5 different zones. The reader may consult Appendix \ref{app1} for the details.

\subsection{PFFs for each planet trap}

We have so far focused on the PFFs summed up for three planet traps: dead zone, ice line and heat transition traps. We now discuss 
the population generated by them separately. Table \ref{table4} summarizes planetary populations produced at three planet traps. 

Many planets formed in dead zone traps become gas giants that eventually fill out Zone 3. This arises from the enhancement of solid densities and 
the efficient formation of gas giants there (see Table \ref{table1}). We find that about half of gas giants piling up beyond 1 AU originate from dead zone traps. 
This is another major result in this paper. It is interesting that a large number of low-mass planets in tight orbits (Zone 5) are also formed 
in dead zone traps. Consequently, dead zone traps have the highest PFFs. This occurs because dead zone traps are effective for the longest time in our models 
(see Table \ref{table1}). In addition, dead zone traps generally distribute planets in the innermost regions of the disk. These two features of dead zone traps both 
enhance planet formation over the entire disk lifetime. 

For ice line traps, we can apply the same discussion as dead zone traps: most planets built there end up in Zone 3. In this case, however, 
the traps become effective only when a certain condition is satisfied (see Table \ref{table1}). In general, the condition cannot be met for the late stage of 
disk evolution. As a result, planet formation in ice line traps is not as efficient as dead zone traps at that time. This leads to the suppression of 
super-Earths and hot Neptunes. The remaining population of gas giants beyond 1 AU (Zone 3) are generated by ice line traps. 

Heat transition traps are the most {\it in}efficient for forming planets. This is explained by the combination of two features of heat transition traps: 
solid densities there are lowest, and they are located in the outermost in protoplanetary disks (see Table \ref{table1}). This leads to the conclusion that 
most planets formed at heat transition traps become low-mass planets in tight orbits. In the fiducial case, half of super-Earths and hot Neptunes are 
generated at heat transition traps. 

In summary, planet traps make different contributions to the populations in these five zones.  This becomes an interesting situation for predicting
the composition of planets.

\begin{table*}
\begin{minipage}{17cm}
\begin{center}
\caption{The results of the fiducial case}
\label{table4}
\begin{tabular}{c|ccccc|c}
\hline
                         &  Zone 1 (\%)   &  Zone 2 (\%)              &  Zone 3 (\%)   &  Zone 4  (\%)   &  Zone 5 (\%)   & Total (\%)    \\ \hline
Dead zone        &  1.1                 &  4.4                            &  12                  &  0                     &  7.1                & 24               \\
Ice line              &  0.32               &  4.6                            &  11                  &  0                     &  0.52              &  16               \\
Heat transition  &  0.21               &  1.6$\times 10^{-3}$  &  1.4                 &  0                     &  6.6                & 8.2                \\    \hline
Total (\%)          &  1.6                 &   9.0                           &   24                 &  0                     &  15                 & 49                \\  \hline
\end{tabular}

The PFFs are given in percentage. 
\end{center}
\end{minipage}
\end{table*}

\section{Parameter study} \label{resu3}

We have so far focused on the results for the fiducial case.  We now explore a larger parameter space in order to discuss 
how prevalent our findings are. More specifically, we examine the model, disk lifetime, and stellar parameters (see Tables \ref{table3} and \ref{table5}).

\begin{table*}
\begin{minipage}{17cm}
\begin{center}
\caption{Summary of a parameter study for the structure of dead zones and the value of $\alpha$}
\label{table5}
\begin{tabular}{c|cccc|ccccc|c}
\hline
Run         &  $\Sigma_{A0}$  & $s_A$  &  $\alpha_A$ &  $\alpha_D$ &  Zone 1 (\%)              &  Zone 2 (\%)              &  Zone 3 (\%)            &  Zone 4 (\%) &  Zone 5 (\%) & Total (\%) \\ \hline
Run A1   &  5                        & 3          &  $10^{-3}$   &  $10^{-4}   $  &  0.1                           & 2.4                             &  28                           & 0                   &  17                & 47            \\  
Run A2   &  50                      & 3          &  $10^{-3}$   &  $10^{-4}   $  &  2.5                           & 6.9                             &  9.9                          & 0                   &  37                & 56            \\  
Run A3   &  20                      & 1.5       &  $10^{-3}$   &  $10^{-4}   $  &  9.6$\times 10^{-3}$  & 2.6                            & 18                            & 0                   &  35                & 55            \\    
Run A4   &  20                      & 6          &  $10^{-3}$   &  $10^{-4}   $  &  2.6                            &  3.5                           &  14                           & 0                   &  35                & 55            \\   \hline
Run B1   &  20                      & 3          &  $10^{-2}$   &  $10^{-4}   $  &  0.21                          &  3.2$\times 10^{-3}$ & 0                              & 0                   &  14                & 14             \\
Run B2   &  20                      & 3          &  $10^{-2}$   &  $10^{-5}   $  &  0.12                          & 4.4                            & 1.6$\times 10^{-3}$ & 0                   & 9.9                & 14             \\
Run B3   &  20                      & 3          &  $10^{-3}$   &  $10^{-5}   $  &  0                               &  0.86                         &  39                           & 0                   &  10                & 50             \\  \hline
\end{tabular}
\end{center}
\end{minipage}
\end{table*}

\subsection{The structure of dead zones}

It is well recognized that dead zones are likely to be an important ingredient for understanding protoplanetary disks and planet formation in the disks 
\citep[e.g.,][HP11, Paper I]{g96,mpt09,zhg10,gnt11}. Nonetheless, it is still uncertain what a "realistic" structure of dead zone is. We have adopted 
a simple analytical model for characterizing the structure of dead zones that is parameterized by the surface density of the active zone ($\Sigma_{A0}$) 
and the power-law index ($s_A$) (see Table \ref{table1}). 

We here explore the effects of both $\Sigma_{A0}$ and $s_A$ on the PFFs by varying these two parameters. Table \ref{table5} tabulates 
our parameter study and the results (see Runs A). We first discuss the effects of $\Sigma_{A0}$, and then examine $s_A$.

We confirmed that the change of $\Sigma_{A0}$ does not change our findings qualitatively (see Runs A1 and A2). In fact, the results show that 
most planets formed at planet traps are most likely to distribute in Zone 3. Also, the coupling of planet traps with the core accretion scenario generates 
a large number of low-mass planets in tight orbits. It is interesting that the population of Zone 3 is a decreasing function of $\Sigma_{A0}$, 
whereas the opposite is established for the population of Zone 5. For the variation of $s_{A}$, we again obtained the results that are qualitatively similar to 
the fiducial case (see Runs A3 and A4). Thus, we can conclude that our findings are valid for a wide range of the structure of dead zones.

\subsection{Disk viscosity - dependence on values of $\alpha$}

In our semi-analytical models, there are two values of $\alpha$ for characterizing the strength of disk turbulence in the dead ($\alpha_D$) and 
active ($\alpha_A$) zones. We discuss how the PFFs are affected by varying them. Table \ref{table5} summarizes our runs and the results (see Runs B). 
We discuss the effects of $\alpha_A$ and $\alpha_D$ separately. 

We first examine $\alpha_A$. The results show that the PFFs of Zone 3 are quite sensitive to the value of $\alpha_A$ (see Runs B1 and B2). 
Since Zone 3 is the favorite target for gas giants in the mass-period diagram, the total PFFs also decrease when $\alpha_A=10^{-2}$. This is indeed expected. 
In general, planet traps are found initially in the outer region of disks where the surface density of disks is determined by $\alpha_A$.  In addition, 
our model dictates that the increment of $\alpha_A$ results in the reduction in the surface density. As a result, the total and Zone 3's PFFs decrease with 
increasing values of $\alpha_A$.  Thus, the cores of most gas giants that end up in Zone 3 are initially formed in the outer region of disks and then transported 
towards the inner region of disks, following the inward movement of planet traps, and accreting as they go.

Compared with $\alpha_A$, the change in the value of $\alpha_D$ does not affect the results very much (compare Fiducial vs Run B3, or Run B1 vs Run B2). 
This is also understood by the same argument as above: the value of $\alpha_D$ is involved with the surface density of dead zones, and planet formation 
takes place mainly beyond the dead zones, so that it does not affect planet formation proceeding in planet traps.

It is interesting that the PFFs of Zone 5 are very insensitive to the variation of $\alpha_A$ and $\alpha_D$. This further supports our findings that 
a considerable number of observed low-mass planets in tight orbits are likely to be planets that cannot accrete gas around them efficiently. 
 
\subsection{The final mass of planets ($f_{fin}$)} \label{para_f_fin}

\begin{figure}
\begin{center}
\includegraphics[width=9cm]{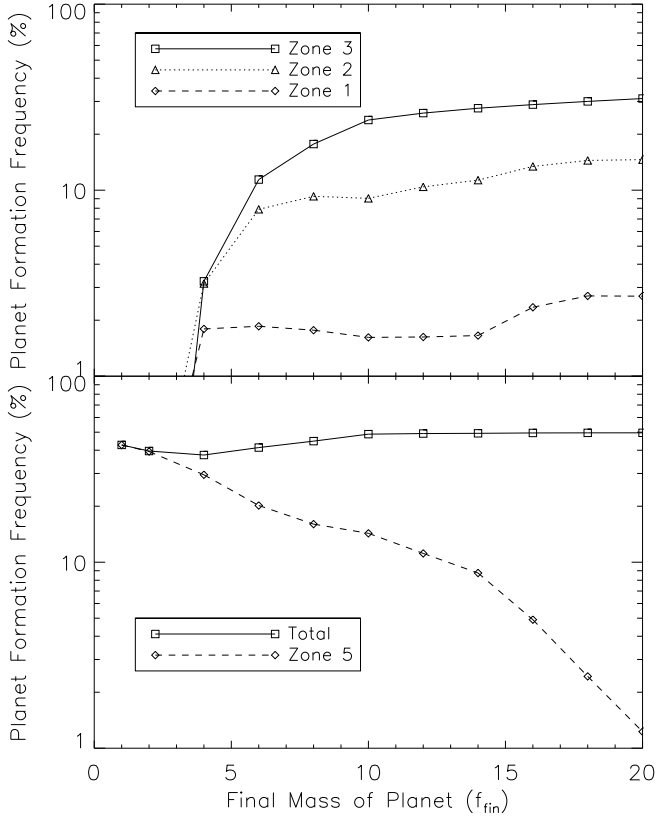}
\caption{The PFFs as a function of $f_{fin}$. The value of $f_{fin}$ is involved with the termination of planetary growth (see equation (\ref{M_max})). 
On the top panel, 
the PFFs of Zones 1, 2, and 3 are denoted by the dashed, the dotted, and the solid lines, respectively, The results show that Zone 3 is a place 
at which most formed gas giants end up for the case that $f_{fin} \ge 5$. On the bottom panel, the total PFFs and those of Zone 5 are shown by the solid 
and dashed lines, respectively. The results show that the total PFFs are not so sensitive to the value of $f_{fin}$. Also, the results suggest that 
the population of low-mass planets within tight orbits is likely to give a clue for putting on a useful constraint on the final stage of gas giant formation. 
Furthermore, the results point out that the entire population of planets fills out Zone 5 if $f_{fin} \le 5$.}
\label{fig4}
\end{center}
\end{figure}

One of the most uncertain parameters is $f_{fin}$, which regulates the termination of planetary growth (see equation (\ref{M_max})). 
Adopting the values given in the fiducial case, we vary only $f_{fin}$ from 1 to 20, in order to examine how important the value of $f_{fin}$ is 
for determining the PFFs of 5 zones. Figure \ref{fig4} shows the results of the PFFs as a function of $f_{fin}$. We find that our results are valid 
for the case that $f_{fin} \ge 5$, and hence most planets formed at planet traps tend to end up in Zone 3 preferentially (see the top panel of Figure \ref{fig4}).  
It is important that the results also show that when $f_{fin} \ge 10$, the PFFs of Zones 1, 2 and 3 are very insensitive to the variation of $f_{fin}$. 
This implies that it is difficult to derive a realistic constraint on the final stage of planet formation from comparing the theoretical calculations with 
the observations of gas giants.

The bottom panel of Figure \ref{fig4} shows the results of the total PFFs and those of Zone 5. It is interesting that the total PFFs do not change very much 
with $f_{fin}$. This indicates that the value of $f_{fin}$ regulates only a fraction of the PFFs in each zone: when a small value of $f_{fin}$ is adopted, 
most formed planets end up in Zone 5 whereas if a large value of $f_{fin}$ is taken, most formed planets fill out Zone 3. This behavior is confirmed 
by the PFFs of Zone 5 which are strongly affected by the change of $f_{fin}$. The results show that the population of Zone 5 is a decreasing 
function of $f_{fin}$. Zone 5 is the most common place for planets to end up if $f_{fin} \la 5$. Thus, we can conclude that a understanding of 
formation mechanisms of super-Earths and hot Neptunes is a key to investigating how planets stop accreting the surrounding gas.

\subsection{Viscous timescales ($\tau_{vis}$)} \label{resu3_1}

It is still not clear how protoplanetary disks evolve with time and dissipate in the final stage of disk evolution \citep[e.g.,][]{a11}. In the models, we assume that 
the long term evolution of disks ($\sim 10^6$ yr) is regulated by the turbulent viscosity and that the end stage of disk evolution is determined by some kind 
of dissipative processes that are represented by exponential functions (see equation (\ref{mdot_exp})). This means that the disk lifetime is defined largely 
by the depletion timescale ($\tau_{dep}$) in our formalism. As a result, we coupled the SPFFs, that are estimated for a given value of $\eta_{dep}$, 
with the weight functions in which the observations of protoplanetary disks are taken into account (see equations (\ref{pfr}) and (\ref{w_pe})).

\begin{figure}
\begin{center}
\includegraphics[width=9cm]{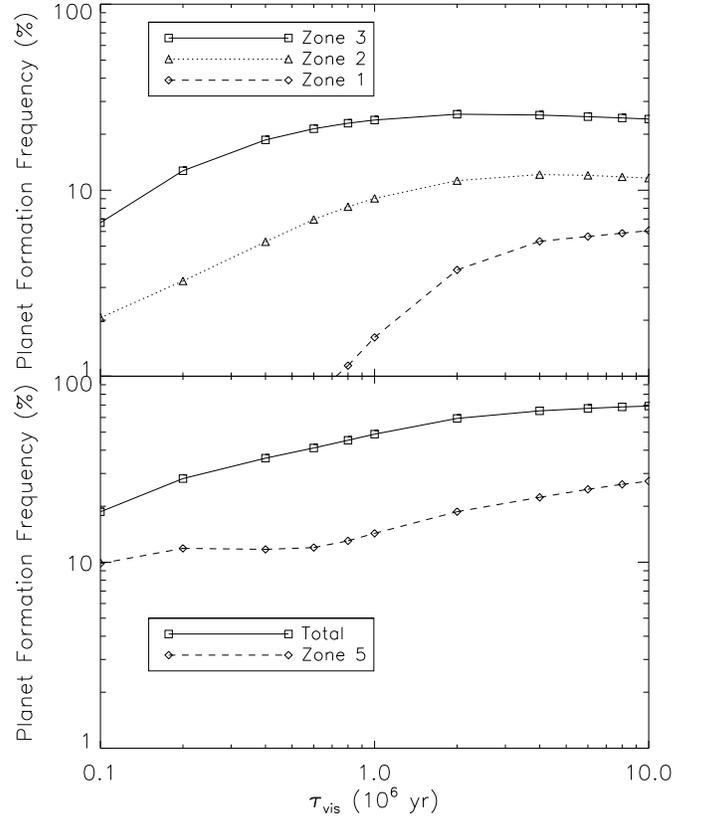}
\caption{The PFFs as a function of $\tau_{vis}$ (as Figure \ref{fig4}). Zone 3 is the best place for gas giants to finally distribute for a variety of $\tau_{vis}$. 
In addition, the results suggest that many low-mass planets in tight orbits (Zone 5) are produced even if the value of $\tau_{vis}$ is varied. 
Thus, our findings are valid for a wide range of $\tau_{vis}$. The results also show that the total PFFs increase with $\tau_{vis}$. This is because 
a large value of $\tau_{vis}$ dictates slow disk evolution, leading to the enhancement of the planet formation efficiency.}
\label{fig5}
\end{center}
\end{figure}

It is interesting to consider the effects of the long term evolution of disks on the PFFs. Figure \ref{fig5} shows the resultant PFFs of 5 zones as a function of 
$\tau_{vis}$, and confirms that our findings given for the fiducial case are regained for a wide range of $\tau_{vis}$. We can therefore conclude that 
when planet traps are coupled with the core accretion scenario, most gas giants tend to fill out Zone 3. In addition, many super-Earths and hot Neptunes are 
also formed due to the same mechanism of forming gas giants.  The only difference between gas giants and low-mass planets arises from the amount of 
gas accretion onto planetary cores.

The results also show that the total PFFs are increasing functions of $\tau_{vis}$. This is a reflection of disk evolution that becomes slow for a large value of 
$\tau_{vis}$. As a result, solid materials become available for a long time and hence there are more chance for planets to form in the disks.

\subsection{Dependence on stellar mass} \label{resu3_2}

Finally, we vary the stellar parameters in order to address how common our findings are for a various type of stars. In our formalism, 
the stellar parameters consist of three valuables: the stellar mass ($M_*$), radius ($R_*$), effective temperature ($T_*$). In order to estimate $R_*$ 
consistently, we adopt the mass-luminosity relation, 
\begin{equation}
\frac{L_*}{L_{\odot}} = \left( \frac{M_*}{M_{\odot}} \right)^a,
\end{equation}
where $L_*$ and $L_{\odot}$ are the stellar and solar luminosities, respectively. It is well known that $a=4$ for $0.5M_{\odot} \leq M_{*} \leq 2 M_{\odot}$. Combining with 
the relation that $L_*/L_{\odot}=(R_*/R_{\odot})^2(T_*/T_{\odot})^4$, the stellar radius becomes a function of both $M_*$ and $T_*$:
\begin{equation}
\frac{R_*}{R_{\odot}}= \left( \frac{M_*}{M_{\odot}} \right)^{a/2} \left( \frac{T_*}{T_{\odot}} \right)^{-2}.
\end{equation}
Table \ref{table6} summarizes the values of $M_*$ and $T_*$.

\begin{figure}
\begin{center}
\includegraphics[width=9cm]{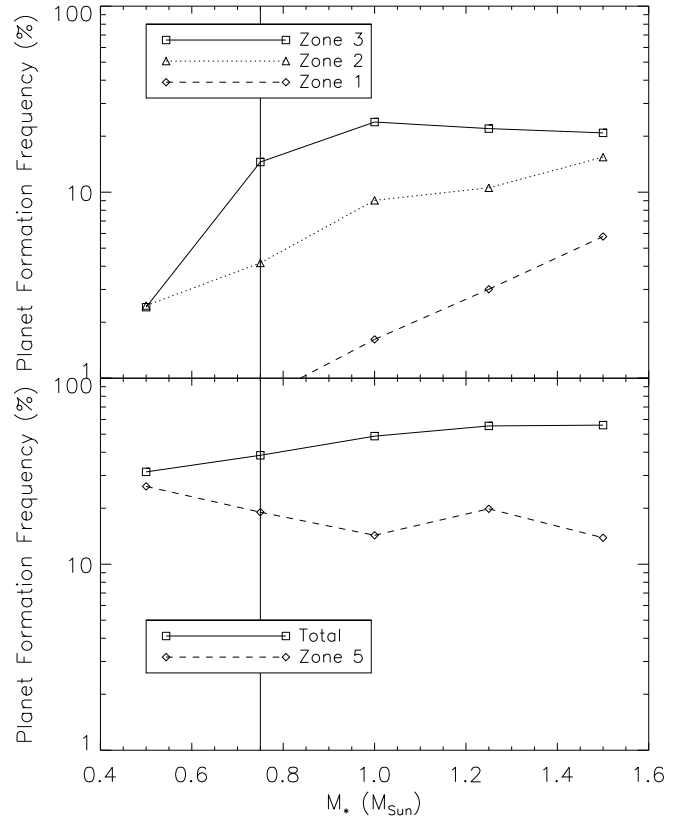}
\caption{The PFFs as a function of stellar mass (as Figure \ref{fig4}). The results confirm that our findings are valid for a various type of stars. 
As the stellar mass increases, the dominant population varies from Zone 5 to Zone 3 (see the vertical solid line), which is consistent with 
the previous studies. In addition, the PFFs of Zone 5 are high for a variety of stellar masses. This indicates that a number of low-mass planets 
in tight orbits are observed for a various type of stars. Also, the results show that the total PFFs increases as the stellar mass increases. 
This is a direct reflection of $\dot{M}$ that is linearly proportional to the total disk mass (see equation (\ref{M_tot})) and is a increasing function of $M_{*}$ 
(see equation (\ref{mdot_exp})).}
\label{fig6}
\end{center}
\end{figure}

Figure \ref{fig6} shows the results of the PFFs for 5 zones as a function of stellar mass.  We find an important result:  the dominant 
population changes from Zone 3 to Zone 5 around 0.75 $M_{\odot}$ (see the vertical solid line). This behavior is consistent with the previous studies which 
show that gas giants are preferentially formed for massive stars \citep[e.g.,][]{il05}. The results also show that the population of Zone 5 is not so sensitive to 
the variation of stellar mass. Thus, we can conclude that super-Earths and hot Neptunes are observed for various types of stars. This trend is now 
being confirmed by both the radial velocity and transit observations \citep[e.g.,][]{mml11,bkb11}. In addition, the results show that the total PFFs are 
an increasing function of $M_{*}$. This is indeed expected in our formalism where the total disk mass, that is linearly proportional to $\dot{M}$, increases 
with $M_{*}$ (see equations (\ref{mdot_exp}) and (\ref{M_tot})). As a result, planet formation proceeds efficiently for massive stars.

In summary, we conclude that most gas giants formed at planet traps tend to distribute beyond 1 AU (Zone 3) for a wide range of disk and stellar parameters. 
More specifically, Zone 3 becomes the preferred target of gas giants for $\Sigma_{A0} \leq 50$, $\alpha_A \leq 10^{-3}$, and $f_{fin} \geq 5 $. For 
any other parameters, the results are qualitatively similar to those of the fiducial case. In addition, our results suggest that many low-mass planets within 
tight orbits (Zone 5) are formed as "failed" cores of gas giants. This is confirmed for a wide range of the disk and stellar parameters. 

\begin{table}
\begin{center}
\caption{Summary of $M_*$ and $T_*$}
\label{table6}
\begin{tabular}{cc}
\hline
$M_*$ ($M_{\odot}$)   &  $T_*$ (K)   \\ \hline
0.5                               &  4000          \\
0.75                             &  4500          \\
1.25                             &  6500          \\
1.5                               &  7500          \\  \hline
\end{tabular}
\end{center}
\end{table}

\section{Discussion} \label{disc}

We here compare our results with the observations of exoplanets, and discuss what the consequences are.


\subsection{Gas giants within 10 AU}

We have shown that the formation of gas giants involves an interesting interplay between the role of planet traps, and the importance of type II migration 
after planets open a gap in their natal disks. At high disk accretion rates (say $\eta_{acc}=10$), the formation efficiency of planetary cores is high 
(since the disk mass is high). As a result, the cores grow relatively quickly, and subsequently become gas giants efficiently. The high disk accretion rates 
also result in planet traps that are initially located far from the host stars. Combining the fact that the efficient formation of gas giants leads to "drop-out" 
of planets from their traps at the early stage of disk accretion, the planets are likely to start undergoing type II migration at $r \simeq 10$ AU. In this case, 
type II migration definitely plays an important role in determining the end points of planets. This is because planets have considerable time over which 
type II migration becomes effective. Thus, the terminal point of planets is regulated largely by $\eta_{dep}$.  For intermediate disk accretion 
rates (say $\eta_{acc}=1$),  planets drop-out from planet traps at $r \simeq 1-5$ AU. This is a combined consequence of less efficient planet formation 
and more effective transport of planets by the traps: the formation of cores of gas giants for this case takes some time, so that they are very likely to 
be transported by the traps for a longer time. This indicates that the remaining time for planets to undergo type II migration is not so long, 
because they leave the traps at the later stage of disk evolution. Nonetheless, it is expected that type II migration still plays a non-neglible role 
since the local viscous timescale, that regulates the type II migration, is relatively short ($\la 10^6$ yr) for this case. Consequently, $\eta_{dep}$ also 
becomes important.

The observations of exoplanets show that the population of close-in planets is not large. For example, \citet{mml11} infer through both CORARIE and 
HARPS observations that the occurrence rate of hot Jupters is less than 1 \%. We note that the population of hot Jupiters in Figure \ref{fig1} is 
significantly exaggerated due to the transit observations. In addition, Figure \ref{fig1} shows that gas giants distributing within $\sim 0.6$ AU are also rare. 

Our results are qualitatively consistent with the observations in a sense that most gas giants tend to end up beyond $\sim 0.6$ AU (see the PFFs in 
Table \ref{table4}). Nonetheless, there is still likely to be a quantitative difference between our results and the observations when the PFFs of Zone 2 
are compared with those of Zone 3. Our results show that the ratio of the PFFs for Zone 3 to Zone 2 is about 2.7 for the fiducial case. This means that 
it is about 2.7 times more likely for gas giants to end up in Zone 3 than Zone 2. The observations, however, imply that the ratio is likely to be higher than 
that value. We estimated that the observed ratio is at least larger than $\sim 4$, by simply counting the number of observed exoplanets distributing in Zone 2 
and 3 (see Figure \ref{fig1}). This suggests that additional physical processes may be needed for enhancing the pile up of gas giants around 1AU more. 
We discuss two possibilities: a slowing mechanism of type II migration and planet formation triggered by planets formed at planet traps.

As discussed in Section \ref{resu1_1}, the populations of Zones 1 and 2 are affected largely by type II migration that transports gas giants formed initially at 
$r \simeq 1-10$ AU toward the vicinity of the central star. The orbital evolution of planets via type II migration is terminated practically when the total disk 
mass becomes very small. Historically, type II migration has not been considered as a problem. This may be because type I migration has been 
more problematic and type II migration is in general slower than type I migration. Nonetheless, the recent population synthesis calculations show 
that too many hot Jupiters are generated and about 90 \% of them must perish somehow in order to reproduce the observations \citep[e.g.,][]{il08v}. 
This may suggest that type II migration is also a problem, so that some kind of slowing mechanisms for the migration may be needed. 

In our calculations, 
type II migration is not so problematic due to two slowing mechanisms: one of them is the presence of dead zones that reduce the local viscous timescale 
by lowering the value of $\alpha$, and the other is the inertia of planets. However, some difference between our results and the observations, that is 
prominent especially in Zone 2, may also support this argument. Since the effectiveness of type II migration is well coupled with the disk evolution 
that is still inconclusive, we will leave a more detail discussion to a future publication \citep{hi13}.
  
The enhancement of the PFFs in Zone 3 could be achieved by subsequent planet formation that is triggered by gas giants formed at planet traps. 
\citet{koi12} have recently investigated the formation of Saturn under the presence of Juipter and shown that once Jupiter is formed somehow and 
opens up a gap in the disk, the formation of a core of Saturn proceeds very rapidly. This can occur due to the pile up of solid materials near 
the pressure maximum that can be formed by the gap. It is important that this mechanism is applied only for the outer edge of the gap, 
since solid materials generally migrate inward due to the so-called head winds and/or the disk accretion \citep[e,g,][]{w77,kl07}. As a result, 
the "second" planet formed in the system distributes beyond the "first" planet. 
In our calculations, it is assumed that planet formation takes place only 
at planet traps in protoplanetary disks, and is obvious that such planets are considered as the "first" planets in the system. This is because, 
without some kind of stopping mechanisms of type I migration, no planets are formed in the system. 
Since such second planets are located beyond the first planets, Zone 3 is more likely to be filled by the second planets. 
 
\subsection{Origins of distant planets}

There is no planet that ends up in Zone 4 in most of the calculations done in this paper (see Sections \ref{resu1_1}, \ref{resu2_1}, and \ref{resu3}, and 
Appendix \ref{app1}). This is indeed in good agreement of the recent deep-imaging surveys that target substellar companions at large orbital 
separations \citep[e.g.,][]{vpm12}. Such dedicated surveys detect only few planets beyond 10 AU. As a result, we can conclude that 
our results are consistent qualitatively with the recent observations of imaged, distant planets.
 
\subsection{Origins of super-Earths and hot Neptunes} \label{disc2}

Both radial velocity and transit observations reveal that the population of super-Earths and hot Neptunes is very large and suggest that 
they are the dominant population \citep[e.g.,][]{hmj10,mml11}.  Our results show that a large fraction of low-mass planets in tight orbits are 
generated due to the same forming mechanism as gas giants. For low-mass planets, the effects of type II migration are much smaller. 
In this regime, what is of  prime importance is that the "failed" cores be transported by the traps to the vicinity of host stars. The resultant PFFs are 
more or less comparable to those of Zone 3 in all the calculations done above. This indicates that a couple of additional mechanisms 
are likely to play a role in forming super-Earths and hot Neptunes. 

Another possibility is that mergers of embryos that essentially take place after gas disks dissipate significantly. This mechanism is identical to 
the one for forming rocky planets like the Earth. There are a number of works that investigate this scenario. For example, \citet{il10} have recently 
performed population synthesis calculations and shown that the population of many super-Earths and hot Neptunes can be produced by mergers of embryos. 

What are implications of this argument? The most important is that different formation mechanisms lead to different compositions of planets. 
In fact, it is expected that when planets are formed as the standard core accretion scenario, the mean density of many super-Earths and hot Neptunes 
may not be as high as that of the Earth, whereas low-mass planets formed via mergers of embryo are likely to results in the high mean density. 
The composition of super-Earths and hot Neptunes has recently received a lot of attention, because they are most likely to be in habitable zones 
(HZs) \citep{kwr93,lfg04,skl07}. In the HZs, the development of life becomes possible due to the presence of liquid water. In order for such planets 
to be literally habitable, it is ideal that they are composed of rocky materials like the Earth. Furthermore, our parameter study done in Section \ref{resu3} 
suggests that the population of low-mass planets in tight orbits is likely to play a key role in examining how planetary growth is terminated 
(see Figure \ref{fig4}). Thus, it is crucial to identify their formation mechanisms and the consequences on their compositions. We emphasize that 
the observational estimate of the composition is now becoming available thanks to the {\it Kepler} mission. 

\section{Conclusions} \label{conc}

We have investigated the statistical properties of planets that are formed in planet traps by changing a number of disk and stellar parameters. 
We have adopted a semi-analytical model developed in Paper I, wherein planet formation and migration are coupled with the time evolution of 
protoplanetary disks that is regulated by both viscous diffusion and dissipative processes (such as photoevaporation). 
The main feature of the model is that rapid type I migration 
is halted at planet traps, and that forming planetary cores subsequently move inwards 
through the evolving disk along with the trap. We have considered three types of planet traps: dead zone, ice line, and 
heat transition traps. We have modified the treatment of a dissipative process, and adopted an "intermediate" behavior of disk evolution.

We begin our analysis by dividing the mass-semimajor axis diagram into 5 zones (see Figure \ref{fig1}). This kind of classification is 
indeed suggested by the observations and our calculations are focused on how these populations arise from our picture of trapped planet formation. 
In order to do this, we have introduced specific planet formation frequencies (SPFFs) (see equation (\ref{spfr})) and (integrated) planet formation 
frequenciess (PFFs) (see equation (\ref{pfr})).   We connect the statistics of our evolutionary tracks - which are governed by two basic parameters
(the disk accretion rate and the disk depletion timescale)  with the data using the PPFs. Our new approach has enabled us to quantitatively estimate 
the contributions of  planet  traps for generating the population of planets in the mass-semimajor axis diagram, as distinct from the 
population synthesis approach. 

We have compared our results with the observations of exoplanets and discussed the implications of our findings. We can conclude that our 
results are qualitatively consistent with the statistics of currently observed exoplanets around solar-type stars. Nonetheless, some additional 
physical processes may be needed for quantitatively reproducing the observations.

We list our major findings below.
 
\begin{enumerate}

\item We have shown that when core accretion-based planet formation is coupled with planet traps, most gas giants end up around 1 AU (Zone 3). 
This is consistent with the observations which show that there is a pile up of gas giants around 1 AU. Also, we have demonstrated that 
a large number of low-mass planets in tight orbits (Zone 5), also known as super-Earths and hot Neptunes, are formed as "failed" cores of gas giants that
form in low mass disks. We summarize the PFFs of the fiducial case here again (also see Table \ref{table4}): PPFs of hot Jupiters are 1.6 \%, 
those of exclusion of gas giants are 9.0 \%, those of gas giants around 1 AU are 24 \%, and those of low-mass planets in tight orbits are 15 \%.

\item We have also shown that planet formation\ in dead zone and ice line traps plays the dominant role in filling out Zone 3, 
whereas the population of super-Earths and hot Neptunes (Zone 5) is generated by planets forming at both dead zone and heat transition traps 
(see Table \ref{table4}). More specifically, both dead zone and ice line traps contribute equally to a pile of gas giants around 1 AU. For Zone 5, 
the contributions from dead zone and heat transition traps are also roughly comparable. 

\item We have performed a parameter study in which four parameters -$\Sigma_{A0}$, $s_A$, $\alpha_A$, and $\alpha_D$ -  are varied (see Table \ref{table5}, 
also see Table \ref{table3} for their definition). These parameters are involved with the structure of dead zones. We have shown that the population of Zone 3 
is very insensitive to the detail structures of dead zones ($\Sigma_{A0}$ and $s_A$). Also, we have demonstrated that the resultant planetary population 
in Zone 3 is affected largely by the disk turbulence in the active region ($\alpha_A$) (rather than the dead zone ($\alpha_D$)). We have shown that, 
in order to reproduce the observational trend, $\alpha_A$ should be comparable or lower than $10^{-3}$. For the population of Zone 5, the change of 
these parameters does not affect the results very much.

\item We have investigated the effects of the final stage of planet formation on the resultant PFFs.  In particular, we have parameterized 
the termination of planetary mass growth by the value of  $f_{fin}$ in equation (\ref{M_max}) (see Figure \ref{fig4}). 
Specifically, we have demonstrated that the PFFs of Zones 1, 2 and 3 are very insensitive to 
the variation of $f_{fin} \geq 5$ whereas the PFFs of Zone 5 vary with $f_{fin}$.  We have also shown that most planets populate Zone 5 for 
$f_{fin} \leq 5$. This suggests that the population of super-Earths and hot Neptunes plays a key role in investigating the final stages of the formation
of massive planets before the surrounding gas disperses entirely.
 
 \item Our results support the need for combined viscous and dissipative effects in disks that end up with the cutoff of the disk accretion rate 
at the final stage of disk evolution. We have found that it is not important how sharply the disk accretion rate is truncated when the distribution of 
the disk lifetime is properly sampled. Thus, our results indicate that the most important parameter for reproducing the observations of exoplanets 
is the disk lifetime.
 
\item We have also examined stellar mass dependency and demonstrated that our findings are achieved for a various type of stars from $0.5 M_{\odot}$ to 
$1.5 M_{\odot}$ (see Figure \ref{fig6}). We have also shown that low-mass planets in tight orbits become dominant in the resultant planetary population 
for low mass stars ($M_{*} \leq 0.7 M_{\odot}$). This is in agreement with the previous studies which show that the formation of gas giants is preferred 
for massive stars. In addition, a large number of super-Earths and hot Neptunes are produced for a wide range of spectral types of stars, 
which is consistent with the observations. 

\item {Finally, we have found that it is very rare to find evolutionary tracks into Zone 4, and we predict that massive planets at large
orbital radii should be rare.   This is in good agreement with recent observational surveys.}

\end{enumerate}

In an accompanying paper, we will examine the effects of disk metallicities and discuss how the PFFs of distinct zones in the mass-semimajor axis diagram 
are affected by them.  


\acknowledgments
The authors thank Tristan Guillot and Shigeru Ida for stimulating discussion that was done during the first stage of this work, 
and an anonymous referee for useful comments on our manuscript. Also, YH thanks the hospitality of Tokyo Institute of Technology for hosting stimulating visits. 
Y.H. was supported by McMaster University, by Graduate Scholarships from the ministry of Ontario (OGS) and the Canadian Astrobiology 
Training Program (CATP), and is currently by EACOA Fellowship that is supported by East Asia Core Observatories Association which consists of 
the Academia Sinica Institute of Astronomy and Astrophysics, the National Astronomical Observatory of Japan, the National Astronomical 
Observatory of China, and the Korea Astronomy and Space Science Institute.  R.E.P. is supported by a Discovery Grant from the Natural 
Sciences and Engineering Research Council (NSERC) of Canada.

\appendix

\section{A: Effects of the evolution of the disk accretion rate on planetary populations} \label{app1}

We here discuss how different treatments of the disk accretion rate affect planetary populations. 

As discussed in Section \ref{model_1}, we have adopted equation (\ref{mdot_exp}) for computing the SPFFs and PFFs. This is because the equation 
may characterize the "mean" behavior of the disk accretion rate that evolves with time. It is interesting to examine how different the resultant planetary 
populations are by adopting different treatments of the disk accretion rate. We consider two cases, one of which is that the disk accretion rate is governed 
only by viscous diffusion (Case 1), the other of which is that the rate is regulated by both viscous diffusion and photoevaporation (Case 2). 

For  pure viscous evolution (Case 1), the disk accretion rate can be written as
\begin{equation}
 \label{mdot_vis2}
 \dot{M}(\tau) \simeq  3 \times 10^{-8} M_{\odot} \mbox{ yr}^{-1} \eta_{acc}   \left( \frac{M_*}{0.5M_{\odot}}\right)^2  
                                  \left(1+ \frac{\tau}{ \tau_{dep}} \right)^{(-t+1)/(t-1/2)}.                          
\end{equation} 
Except for a term of the exponential function, this is identical to equation (\ref{mdot_exp}). In order to calculate the PFFs for 5 zones 
(as done in the above sections), we have replaced $\tau_{vis}$ with $\tau_{dep}$. As discussed in Section \ref{model_1}, 
this evolution gives a long tail in the disk accretion rate, so that the disk lifetimes are very long (see Figure \ref{fig2}). 

For sharply truncated photoevaporation (Case 2), we adopt the following equation:
\begin{equation}
 \label{mdot_pe2}
 \dot{M}(\tau) \simeq  3 \times 10^{-8} M_{\odot} \mbox{ yr}^{-1} \eta_{acc}   \left( \frac{M_*}{0.5M_{\odot}}\right)^2  
                               \left[ 1 -  \left(  A  + \tanh \left( \frac{\tau - \tau_{dep}}{\tau_{pe,trans}} \right) \right) \right]
                            \left(1+ \frac{\tau}{ \tau_{vis}} \right)^{(-t+1)/(t-1/2)},                          
\end{equation} 
where $A=| \tanh [ (\tau_{int} - \tau_{dep})/(\tau_{pe,trans}) ] |$. We have adopted $\tanh$-functions for representing the two-timescale nature of disk 
clearing. The timescale $\tau_{pe,trans}=10^5$ yr determines how fast such disk clearing occurs. As shown in Figure \ref{fig2} (see the dashed line), 
the evolution of the disk accretion rate expressed by equation (\ref{mdot_pe2}) provides a rapid decline, so that the disk lifetime can be well defined for 
this case. This is consistent well with the results of the more detailed numerical simulations of disks that are both affected by viscous diffusion and 
photoevaporation \citep[e.g.,][]{gdh09,oec11}.

Adopting these disk accretion rates, we compute the PFFs as done in the above sections. We perform 2 kinds of runs: in one of them, 
we use the same parameters as the fiducial case in which  we set $\mu_{lifetime}=1.5$ and $\sigma_{lifetime}=3$ 
($w_{lifetime}$, see equation (\ref{w_pe})). For the other kind of runs, we change the value of $\mu_{lifetime}$ and $\sigma_{lifetime}$ 
in the weight function ($w_{lifetime}$), so that we set $\mu_{lifetime}=3$ and $\sigma_{lifetime}=8$. 
We discuss the results separately for these two kinds of runs.

For the fiducial case, Tables \ref{tableA.1} and \ref{tableA.2} summarizes the results of Case 1 and 2, respectively. 
The results can be understood by the argument given in Section \ref{resu1_1}: long-lived disks are needed for filling out Zones 1 and 5. 
On the other hand, Zone 3 are occupied for a wide range of disk parameters. As a result, Case 1 that extends the disk lifetimes enhances 
the PFFs of Zones 1 and 5 significantly, while Case 2 that sharply truncates the disk evolution puts planets predominately in Zone 3. 
It is interesting that both cases generate some planets that end up in Zone 4, although the resultant PFFs are very small.  

Now consider the effect of broadening the range of disk accretion times and lengthening the median lifetime.
This is accomplished by an additional parameter study in which the values of $\mu_{lifetime}$ and $\sigma_{lifetime}$ are varied. 
 
As discussed in Section \ref{weight_func}, our choices of $\mu_{lifetime}$ and $\sigma_{lifetime}$ are reasonable in a sense that 
the resultant existence probability of disks well reproduces the observational results in which the disk fraction is fitted well by the exponential function 
with the e-folding timescale 2.5 Myr. The recent observations of disks, however, also imply that the median disk lifetime is likely to be $\sim$ 3 Myr with 
substantial long tails toward older ages. Thus, it may be more appropriate to treat long-lived disks more properly, especially 
when the sharp photoevaporative truncation is considered. This is because the truncation leads to the well-defined disk lifetime. 
Thus, we change the value of $\mu_{lifetime}$ from 1.5 to 3, and those of $\sigma_{lifetime}$ from 3 to 8, 
in order to take long-lived disks into account more appropriately. Since we now attempt to accurately estimate the PFFs for long-lived disks, 
we also vary the range of $\eta_{dep}$  from $0.1 < \eta_{dep} < 10$ to $1 < \eta_{dep} < 100$. The former range is used in all the calculations 
that are discussed  in the above sections. We confirmed that there is almost no contribution to the PFFs from the range ($0.1 < \eta_{dep} < 1$) in this case, 
so that the shift in the range of $\eta_{dep}$ does not affect our results. 

We examine only Case 2 in this situation. Table \ref{tableA.3} tabulates the results. The results show that 
the resultant PFFs are qualitatively consistent with the fiducial case (see Table \ref{table4}), although there are some quantitative differences. Thus, 
even when the sharp photoevaporative truncation is adopted for the evolution of the disk accretion rate, we can generate the observational trends 
of exoplanets: most formed gas giants tend to distribute around 1 AU (Zone 3) with fewer populations of close-in massive planets (Zones 1 and 2). 
The PFFs of Zone 5 are the second highest in the calculation.

In summary, we conclude that the shape of the disk accretion rate, especially at the end stage of disk evolution, is not so important in the resultant PFFs. 
The most important parameter is the disk lifetime in order to reproduce the observational features of exoplanets.

\begin{table*}
\begin{minipage}{17cm}
\begin{center}
\caption{The results of Case 1 (viscous diffusion only)}
\label{tableA.1}
\begin{tabular}{c|ccccc|c}
\hline
                         &  Zone 1 (\%)   &  Zone 2 (\%)              &  Zone 3 (\%)              &  Zone 4  (\%)         &  Zone 5 (\%)   & Total (\%)    \\ \hline
Dead zone        &  10.2               &  8.5$\times 10^{-2}$  &  5.5$\times 10^{-2}$ &  0                            &  18.8               & 29               \\
Ice line              &  11.5               &  0                               &  0                              &  0                            &  3.7                 & 15.2               \\
Heat transition  &  17                  &  6.3$\times 10^{-2}$  &  2.8$\times 10^{-2}$ & 2.6$\times 10^{-3}$ &  10.1              & 27.2                \\    \hline
Total (\%)          &  38.7               &  14.3$\times 10^{-2}$&  8.3$\times 10^{-2}$ & 2.6$\times 10^{-3}$ &  32.6              & 71.4                \\  \hline
\end{tabular}
\end{center}
\end{minipage}
\end{table*}

\begin{table*}
\begin{minipage}{17cm}
\begin{center}
\caption{The results of Case 2 (viscous diffusion and photoevaporation)}
\label{tableA.2}
\begin{tabular}{c|ccccc|c}
\hline
                         &  Zone 1 (\%)   &  Zone 2 (\%)   &  Zone 3 (\%)   &  Zone 4  (\%)             &  Zone 5 (\%)                & Total (\%)     \\ \hline
Dead zone        &  0                   &  0                    &  8.1                 &  0                               &  3.1$\times 10^{-2}$    & 8.1               \\
Ice line              &  0                   &  0                    &  9.1                 &  5.1$\times 10^{-2}$  &  1.3                              & 10.4             \\
Heat transition  &  0                    &  0                   &  0                    &  0                               &  0                                 & 0                  \\    \hline
Total (\%)          &  0                    &  0                   &  17.2               &  5.1$\times 10^{-2}$  &  1.3                              & 18.5             \\  \hline
\end{tabular}
\end{center}
\end{minipage}
\end{table*}

\begin{table*}
\begin{minipage}{17cm}
\begin{center}
\caption{The results of Case 2 for the case that $\mu_{lifetime}=3$ and $\sigma_{lifetime}=8$}
\label{tableA.3}
\begin{tabular}{c|ccccc|c}
\hline
                         &  Zone 1 (\%)              &  Zone 2 (\%)            &  Zone 3 (\%)   &  Zone 4  (\%)             &  Zone 5 (\%)                & Total (\%)     \\ \hline
Dead zone        &  0.23                         &  0.2                          &  19.9               &  0                               &  2.0                             & 22.3              \\
Ice line              &  2.2$\times 10^{-2}$ &  0.41                        &  19.3               &  0.16                          &  0.77                           & 20.7              \\
Heat transition  &  1.7$\times 10^{-2}$ &  2.2$\times 10^{-2}$ &  0.97              & $\sim$0                      &  0.15                           & 1.2                \\    \hline
Total (\%)          &  0.27                         &  0.63                         &  40.2              &  0.16                          &  2.92                           & 44.2              \\  \hline
\end{tabular}
\end{center}
\end{minipage}
\end{table*}






\bibliographystyle{apj}

\bibliography{apj-jour,adsbibliography}    

\end{document}